\pdfoutput=1
\documentclass[sigconf]{acmart}

\AtBeginDocument{%
  \providecommand\BibTeX{{%
    \normalfont B\kern-0.5em{\scshape i\kern-0.25em b}\kern-0.8em\TeX}}}
    
\usepackage{xspace}

\usepackage{algorithm}
\usepackage[noend]{algpseudocode}

\usepackage{tikz}
\newcommand{\circlednumber}[1]{\raisebox{0.5pt}{\protect%
  \tikz[baseline=(myanchor.base)]{
  \node[circle,fill=.,inner sep=1pt,minimum size=4mm] (myanchor) {\color{-.}\footnotesize #1};}%
}}

\newif\ifcameraready
\camerareadytrue

\definecolor{amber}{rgb}{1.0, 0.49, 0.0}
\definecolor{darkgreen}{rgb}{0.0, 0.2, 0.13}
\definecolor{darkbyzantium}{rgb}{0.36, 0.22, 0.33}
\definecolor{darkseagreen}{rgb}{0.56, 0.74, 0.56}
\definecolor{darkspringgreen}{rgb}{0.09, 0.45, 0.27}
\definecolor{dollarbill}{rgb}{0.52, 0.73, 0.4}

\definecolor{darkcyan}{rgb}{0.0, 0.55, 0.55}
\definecolor{forestgreen}{rgb}{0.0, 0.27, 0.13}
\definecolor{azure}{rgb}{0.0, 0.5, 1.0}

\definecolor{darkpink}{rgb}{0.88, 0.28, 0.54}

\newcommand{\hl}[1]{{\color{black}#1}}

\newcommand{\sg}[1]{{\color{black}#1}}
\newcommand{\nour}[1]{{\color{black}#1}}

\newcommand{\revonur}[1]{{\color{black}#1}}

\newcommand{\revIII}[1]{{\color{black}#1}}
\newcommand{\comm}[1]{{\color{NavyBlue}#1}}

\newcommand{\revIV}[1]{{\color{black}#1}}
\newcommand{\revV}[1]{{\color{black}#1}}

\newcommand{\zulalI}[1]{{\color{black}#1}}
\newcommand{\revonurcan}[1]{{\color{black}#1}}

\newcommand{\lijoel}[1]{{\color{black}#1}}

\definecolor{mypink}{RGB}{224,8,95}
\definecolor{myblue}{RGB}{31,116,186}
\definecolor{mygreen}{RGB}{35,155,51}
\definecolor{myorange}{RGB}{201,108, 32}
\definecolor{mypurple}{RGB}{122,32,201}

\definecolor{dblue}{rgb}{0.00, 0.00, 0.55}
\definecolor{ddblue}{rgb}{0.00, 0.00, 0.90}

\definecolor{magenta}{rgb}{255,0,255}
\definecolor{red}{rgb}{255,0,0}
\newcommand{\gagan}[1]{{\color{black}#1}}

\newcommand{\damla}[1]{{\color{black}#1}}
\newcommand{\sgh}[1]{{\color{black}#1}}
\newcommand{\sghi}[1]{{\color{black}#1}}

\newcommand{\mech}{SeGraM\xspace} 
\newcommand{\ba}{BitAlign\xspace}
\newcommand{\ms}{MinSeed\xspace}

\newcommand{\damlaI}[1]{{\color{black}#1}}
\newcommand{\damlaISCA}[1]{{\color{black}#1}}
\newcommand{\sgis}[1]{{\color{black}#1}}

\newcommand{\revISCA}[1]{{\color{black}#1}}

\newcommand{\reviscaI}[1]{{\color{black}#1}}
\newcommand{\reviscaII}[1]{{\color{black}#1}}
\newcommand{\reviscaIII}[1]{{\color{black}#1}}
\newcommand{\reviscaIV}[1]{{\color{black}#1}}
\newcommand{\reviscaV}[1]{{\color{black}#1}}

\newcommand{\sgc}[1]{{\color{black}#1}}
\newcommand{\sgd}[1]{{\color{black}#1}}

\newcommand{\circleblack}[1]{{\color{black}#1}}

\usepackage[binary-units=true]{siunitx}

\usepackage[us,12hr]{datetime}
\newcommand{\versionnum}[0]{4.3 ~--~ \today ~--~ \currenttime \xspace EDT}

\usepackage{enumitem}

\DeclareSIUnit{\nothing}{\relax}
\DeclareSIUnit{\basepair}{bp}

\usepackage{setspace}

\copyrightyear{2022} 
\acmYear{2022} 
\setcopyright{acmlicensed}\acmConference[ISCA '22]{The 49th Annual International Symposium on Computer Architecture}{June 18--22, 2022}{New York, NY, USA}
\acmBooktitle{The 49th Annual International Symposium on Computer Architecture (ISCA '22), June 18--22, 2022, New York, NY, USA}
\acmPrice{15.00}
\acmDOI{10.1145/3470496.3527436}
\acmISBN{978-1-4503-8610-4/22/06}

\newcommand{\affilBNG}{$^{1}$}
\newcommand{\affilETH}{$^{2}$}
\newcommand{\affilBilkent}{$^{3}$}
\newcommand{\affilIntel}{$^{4}$}
\newcommand{\affilCMU}{$^{5}$}
\newcommand{\affilUIUC}{$^{6}$}

\begin{document}

\setstretch{0.96}

\title{\mech: A Universal Hardware Accelerator for \\ Genomic Sequence-to-Graph and Sequence-to-Sequence Mapping}

\author{
{{Damla Senol Cali\affilBNG$^{,}$\affilETH}\quad%
{Konstantinos Kanellopoulos\affilETH}\quad%
{Jo\"el Lindegger\affilETH}\quad%
{Z{\"u}lal Bing{\"o}l\affilBilkent}}\\%
{{Gurpreet S. Kalsi\affilIntel}\quad
{Ziyi Zuo\affilCMU}\quad%
{Can Firtina\affilETH}\quad%
{Meryem Banu Cavlak\affilETH}\quad
{Jeremie Kim\affilETH}}\\
{{Nika Mansouri Ghiasi\affilETH}\quad
{Gagandeep Singh\affilETH}\quad
{Juan Gómez-Luna\affilETH}\quad
{Nour Almadhoun Alserr\affilETH}}\\
{{Mohammed Alser\affilETH}\quad
{Sreenivas Subramoney\affilIntel}\quad%
{Can Alkan\affilBilkent}\quad%
{Saugata Ghose\affilUIUC}\quad%
{Onur Mutlu\affilETH}}\vspace{5pt}\\%
{\Large \affilBNG Bionano Genomics \quad \affilETH ETH Z{\"u}rich \quad \affilBilkent Bilkent University \quad \affilIntel Intel Labs}\\
{\Large \affilCMU Carnegie Mellon University \quad \affilUIUC University of Illinois Urbana-Champaign}
}

\renewcommand{\authors}{Damla Senol Cali, Konstantinos Kanellopoulos, Joel Lindegger, Zülal Bingöl, Gurpreet S. Kalsi, Ziyi Zuo, Can Firtina, Meryem Banu Cavlak, Jeremie Kim, Nika Mansouri Ghiasi, Gagandeep Singh,  Juan Gómez-Luna, Nour Almadhoun Alserr, Mohammed Alser, Sreenivas Subramoney, Can Alkan, Saugata Ghose, and Onur Mutlu}

\renewcommand{\shortauthors}{D. Senol Cali, et al.}
\renewcommand{\shorttitle}{SeGraM: A Universal Hardware Accelerator for Genomic Sequence-to-Graph and Sequence-to-Sequence Mapping}

\begin{abstract}

\damlaISCA{\sgis{A critical step of genome sequence analysis is the \emph{mapping} of sequenced DNA fragments (i.e., \emph{reads}) collected from an individual to a known linear reference genome sequence (i.e., \emph{sequence-to-sequence mapping}).}
\damlaISCA{Recent works replace the linear \reviscaII{reference} sequence with a \reviscaII{graph-based representation of the reference genome}, which captures the 
genetic variations and diversity across many individuals 
in a population. \reviscaIII{Mapping} \reviscaII{reads to the graph-based reference genome (i.e., \emph{sequence-to-graph mapping}) results in notable quality improvements in genome analysis.}}}
\reviscaII{Unfortunately, while}
sequence-to-sequence mapping is well studied with many available tools \reviscaII{and accelerators}, sequence-to-graph mapping is a more difficult computational problem, with \reviscaII{a \reviscaIII{much smaller} number of practical software tools currently available.}

We analyze \reviscaII{two} state-of-the-art sequence-to-graph mapping tools and reveal four key issues.
\reviscaII{
\reviscaIII{We find that} there is a pressing need to have a specialized, high-performance, scalable, and low-cost \reviscaIV{algorithm/hardware} co-design that alleviates bottlenecks in both the seeding and alignment steps of sequence-to-graph mapping. Since sequence-to-sequence mapping can be treated as a special case of sequence-to-graph mapping, we aim to design an accelerator that is efficient for \emph{both} linear \emph{and} graph-based read mapping.}

To this end, we propose \mech, 
\reviscaII{a \emph{universal \reviscaIII{algorithm/hardware} co-designed genomic mapping accelerator} that can \reviscaIII{effectively and efficiently} support both \underline{se}quence-to-\underline{gra}ph \underline{m}apping and sequence-to-sequence mapping, for both short and long reads. 
To our knowledge, \mech is the first algorithm/hardware co-design for accelerating sequence-to-graph mapping.}
\mech consists of two main components: (1)~\ms, the \emph{first} \underline{min}imizer-based \underline{seed}ing accelerator, which finds the candidate locations in a given genome graph; and (2)~\ba, the \emph{first} \underline{bit}vector-based sequence-to-graph \underline{align}ment accelerator, which performs alignment between a given read and the subgraph identified by MinSeed. We couple \mech with high-bandwidth memory to exploit low latency and \reviscaI{highly-parallel memory access, which alleviates the memory bottleneck.} 

\reviscaII{We demonstrate that \mech provides significant improvements for multiple steps of the sequence-to-graph (i.e., S2G) and sequence-to-sequence (i.e., S2S) mapping pipelines. First,
\mech outperforms state-of-the-art S2G mapping tools by $5.9\times$/$3.9\times$ and $106\times$/\-$742\times$ for long and short reads, respectively, while reducing power consumption by $4.1\times$/$4.4\times$ and $3.0\times$/$3.2\times$. Second, \ba outperforms \reviscaIII{a state-of-the-art} S2G alignment tool by $41\times$--$539\times$ and \reviscaIII{three} S2S alignment accelerators by $1.2\times$--$4.8\times$. 
We conclude that \mech is a high-performance and low-cost universal genomics mapping accelerator that efficiently supports both sequence-to-graph and sequence-to-sequence mapping pipelines.}

\end{abstract}



\maketitle

\section{Introduction} \label{sec:introduction}

Genome sequencing, the process used to determine the DNA sequence of an organism, has led to \nour{many} notable advancements in \nour{several fields, such as} personalized medicine \reviscaII{(e.g.,~\cite{alkan2009personalized, flores2013p4,ginsburg2009genomic,chin2011cancer,Ashley2016,chan2011personalized,offit2011personalized})}, \nour{outbreak tracing} \reviscaII{(e.g.,~\cite{wang2020initial,nikolayevskyy2016whole,qiu2015whole,gilchrist2015whole,quick2016real,houldcroft2017clinical,guo2020origin})}, evolutionary biology \reviscaII{(e.g.,~\cite{ellegren2014genome,Prado-Martinez2013,Prohaska2019,hudson2008sequencing})}, and forensic science \reviscaII{(e.g.,~\cite{yang2014application,borsting2015next,alvarez2017next,berglund2011next})}.
Contemporary genome sequencing machines are unable to determine the \emph{base pairs} (i.e., A, C, G, T nucleobases) of the \reviscaII{entire} DNA sequence.
Instead, the machines take a DNA sequence and break it down into small fragments, \reviscaII{called} \emph{reads}, whose base pairs can be \reviscaII{reasonably accurately} identified.
As an example, human DNA consists of approximately \reviscaII{3.2} billion base pairs, while reads, depending on the sequencing technology, range in size from a few hundred~\reviscaII{\cite{hu2021next,iseqwebpage,miniseqwebpage,miseqwebpage,nextseqwebpage,novaseqwebpage}} to a few million~\reviscaII{\cite{cali2017nanopore,amarasinghe2020opportunities,jain2018nanopore,hu2021next,minionwebpage,gridionwebpage,promethionwebpage,sequelwebpage}} base pairs.  
Computers then reconstruct the reads back into a full DNA sequence.
In order to find the locations of the reads in the correct order, \emph{read-to-reference} mapping is performed. The reads are \emph{mapped} to a \emph{reference genome} (i.e., a complete representative DNA sequence of a particular organism) for the same species.

\reviscaII{A single \sgis{(i.e., a linear)} reference genome is not representative of different sets of individuals (e.g., subgroups) in a species and using a single reference genome for an entire species may \emph{bias} the mapping process (i.e., \emph{reference bias}) due to the \emph{genetic diversity} \lijoel{that} exists within a population~\cite{garrison2018variation,martiniano2020removing,chen2021reference,jain2021variant,vaddadi2019read,ballouz2019time,paten2017genome,yang2019one}}. 
For example, the African genome, \reviscaIII{with all known genetic variations within the populations of African descent,} contains 10\% more DNA bases than the current \reviscaII{linear} human reference genome~\cite{sherman2019assembly}.
Combined with errors that can be introduced during genome sequencing (with error rates as high as 5--10\% for long reads~\cite{jain2018nanopore, weirather2017comprehensive,ardui2018single,van2018third,cali2017nanopore,hu2021next,amarasinghe2020opportunities}),
reference bias can lead to significant inaccuracies during mapping.
\sghi{This can create \nour{many} issues for a wide range of genomic studies, from identifying mutations that lead to cancer~\cite{computational2018computational}, to tracking mutating variants of viruses such as SARS-CoV-2~\cite{computonics2020hackathon}, where \reviscaII{detecting the variations that exist in the sequenced genome accurately
is of critical importance for both diagnosis and treatment ~\cite{ambler2019gengraph}.}}

\begin{figure}[h!]
\centering
\vspace{-8pt}
\includegraphics[width=0.85\columnwidth,keepaspectratio]{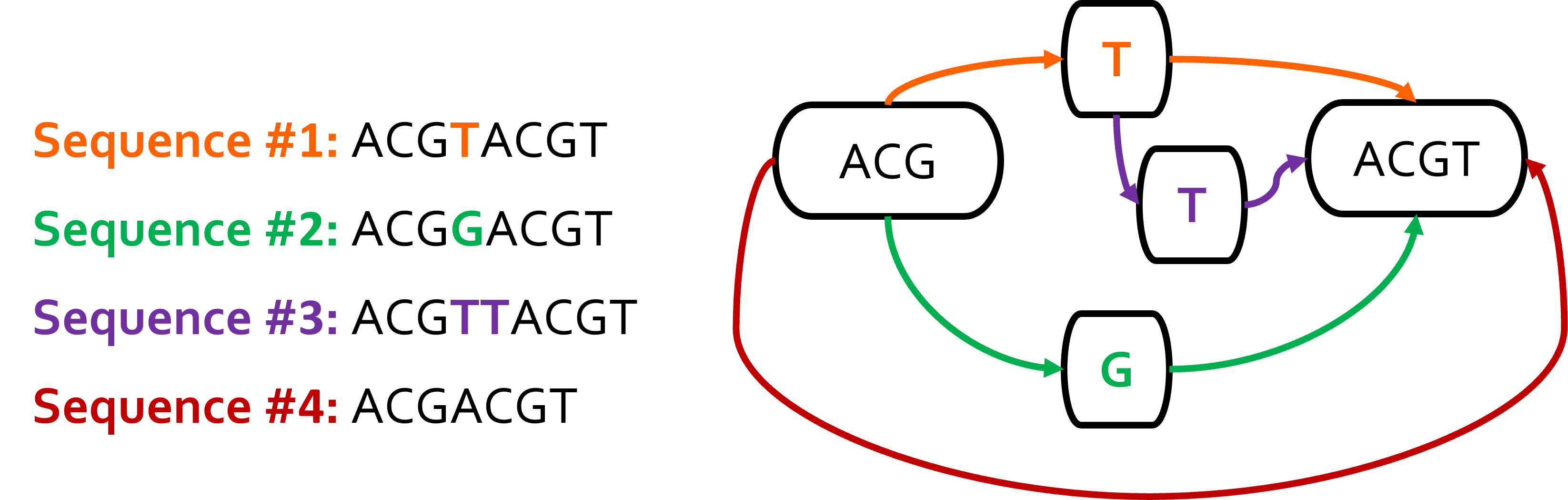}
\vspace{-8pt}
\caption{\reviscaII{Example of a genome graph that represents 4 related but different genomic sequences.}} 
\label{fig:genomegraph-example}
\vspace{-6pt}
\end{figure}

An \sgis{increasingly popular} technique to overcome reference bias is the use of graph-based representations of a species' genome, known as \emph{genome graphs}~\reviscaII{\cite{pevzner2001eulerian,paten2017genome,ameur2019goodbye,kaye2021genome,rakocevic2019fast,eizenga2020pangenome}}.
A genome graph \reviscaII{enables a compact representation of the linear reference genome, \emph{combined with} the}
\emph{known genetic variations} in the \nour{entire} population as a graph-based data structure.
\damlaISCA{As we show in Figure~\ref{fig:genomegraph-example},} a node represents one or more base pairs, \reviscaIV{an edge enables two nodes to be connected to each other, and base pairs in connected nodes represent the sequence of base pairs in the \reviscaV{genomic} sequence.}
Multiple outgoing directed edges from a node captures genetic variations.

\reviscaII{Genome graphs are growing in popularity for a number of genomic applications,
such as (1)~variant calling~\cite{garrison2018variation,rakocevic2019fast,olson2021precisionfda}, which identifies the genomic differences between the sequenced genome and the reference genome; (2)~genome assembly~\cite{compeau2011apply, pevzner2001eulerian, zerbino2008velvet, simpson2009abyss}, which reconstructs the entire sequenced genome using the reads without utilizing a known reference genome sequence; (3)~error correction~\cite{salmela2014lordec,rautiainen2020graphaligner,zhang2020comprehensive}, which corrects the noisy regions in long reads due to sequencing errors; and (4)~multiple sequence alignment~\cite{paten2011cactus, lee2002multiple,li2020design}, which aligns three or more biological sequences of similar length.} \damlaISCA{With \sgis{the} increasing importance and usage of genome graphs, having \reviscaII{fast and efficient techniques and tools} for mapping genomic sequences to \reviscaII{genome} graphs \sgis{is now} crucial.}


\nour{
\sgis{Compared to} sequence-to-sequence mapping, \sgis{where} an organism's reads are mapped to the \reviscaV{single linear reference genome,} 
sequence-to-graph mapping}
captures the inherent genetic diversity \reviscaII{within} a population. \nour{This} results in significantly more accurate \nour{read-to-reference mapping}~\cite{garrison2018variation,rautiainen2020graphaligner,li2020design,yang2019one,kim2019graph}. \reviscaII{For example, sequenced reads from samples that are not \reviscaIII{represented} in the samples used for constructing the reference genome may not align at all or incorrectly align when they originate from a region that differs from the reference genome. This can result in \reviscaIII{failure} to detect disease-related genetic variants. \reviscaIII{However, if (1)~we incorporate the known disease-related genetic variants in our read mapping process using a genome graph and (2)~the sequenced sample contains one or \reviscaIV{more} of these variants, we can accurately detect the \reviscaIV{variant(s).}}}

\begin{figure}[h!]
\centering
\vspace{-6pt}
\includegraphics[width=\columnwidth,keepaspectratio]{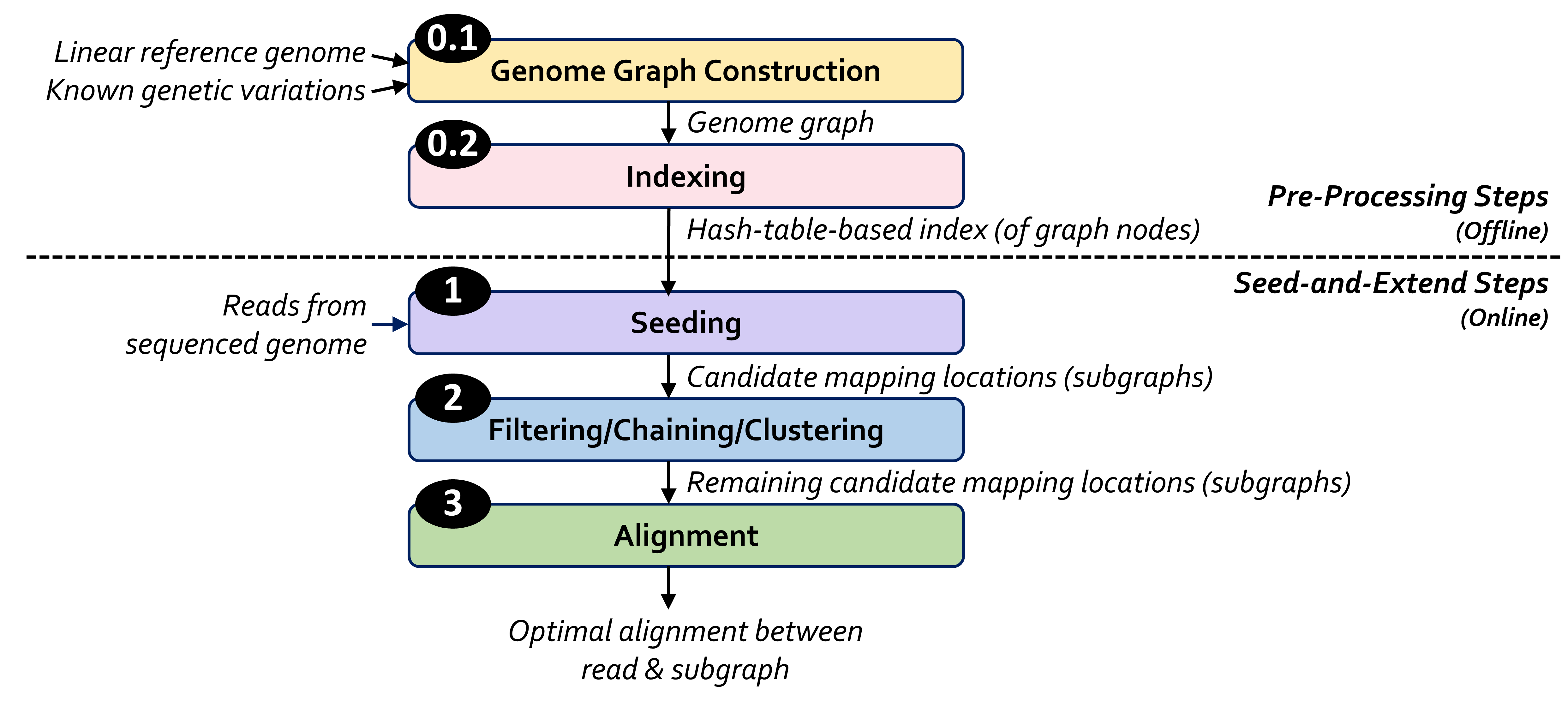}
\vspace{-20pt}
\caption{\reviscaII{Sequence-to-graph mapping pipeline.}} 
\label{fig:mapping-pipeline}
\vspace{-6pt}
\end{figure}

Figure~\ref{fig:mapping-pipeline} shows the sequence-to-graph mapping pipeline, which
follows the \emph{seed-and-extend strategy}~\cite{garrison2018variation,rautiainen2020graphaligner}, similar to sequence-to-sequence mapping~\cite{alser2020accelerating}. \reviscaIV{
The pipeline is preceded by two offline pre-processing steps. The first offline pre-processing step constructs the genome graph using a linear reference genome and a set of known variations~\circlednumber{0.1}. The second offline pre-processing step indexes the nodes of the graph and generates a hash-table-based index~\circlednumber{0.2} for fast lookup. 
When reads from a sequenced genome are received, the pipeline tries to map them to the pre-processed reference graph using three online steps. First, the \emph{seeding} step~\circlednumber{\circleblack{.}1\circleblack{.}} is executed, where each read is fragmented into sub-strings (called \emph{seeds}) and exact matching locations of these seeds (i.e., candidate mapping locations) are found within the graph nodes using the index. Second, the optional \emph{filtering}, \emph{chaining}, or \emph{clustering} step~\circlednumber{\circleblack{.}2\circleblack{.}} is performed to decrease the number of required alignments in the next step. Third, the \emph{alignment} step~\circlednumber{\circleblack{.}3\circleblack{.}} is performed between all 
remaining candidate mapping locations (i.e., subgraphs) within the graph and the query read to find the optimal alignment.
}

Prior works\nour{~\cite{turakhia2018darwin,cali2020genasm, fujiki2018genax,alser2021technology,alser2017gatekeeper,alser2020sneakysnake,kim2019airlift,alser2020accelerating,kim2018grim}} \reviscaIII{show} that read-to-reference mapping is one of the major bottlenecks of the full genome sequence analysis pipeline, and that it can benefit from \reviscaII{\reviscaIV{algorithm/hardware} co-design~\cite{alser2020accelerating,singh2021fpga}} that takes advantage of specialized hardware accelerators.
Given the additional complexities and overheads of processing a genome graph instead of a linear reference genome, graph-based analysis exacerbates \sgis{the bottlenecks of} read-to-reference mapping. Due to the nascent nature of sequence-to-graph mapping, \reviscaII{\reviscaIII{a much smaller number of
software tools (and no hardware accelerators)} exist for sequence-to-graph mapping~\cite{rautiainen2020graphaligner, garrison2018variation, rakocevic2019fast, kim2019graph, rautiainen2019bit, ivanov2020astarix, jain2020complexity,kavya2019sequence,darby2020vargas,chang2020distance,ivanov2022fast} compared to the traditional sequence-to-sequence mapping.} 

\reviscaI{
In order to \reviscaII{identify and quantify} the performance bottlenecks of existing tools, we analyze
GraphAligner~\cite{rautiainen2020graphaligner} and vg~\cite{garrison2018variation}, \reviscaII{two} state-of-the-art software \reviscaII{tools} for sequence-to-graph mapping. Based on our analysis \reviscaII{(Section~\ref{sec:motivation})}, we make four key observations. 
\reviscaIII{
(1)~Among the three \reviscaIV{online} steps of the read mapping pipeline (i.e., \emph{seeding}, \emph{filtering}, and \emph{alignment}), sequence-to-graph alignment i)~constitutes 50--95\% of the end-to-end execution of sequence-to-graph mapping, and ii)~is even more expensive than its counterpart in the traditional read mapping pipeline~\cite{turakhia2018darwin, fujiki2018genax, cali2020genasm} since a graph-based representation of the genome is more complex to process (both computationally and memory-wise) than the linear representation. 
(2)~Alignment suffers from high cache miss rates, due to the high amount of internal data that is generated and reused during this step. 
(3)~Seeding suffers from \reviscaIV{the main memory (DRAM) latency bottleneck}, due to the high number of irregular memory accesses generated when querying the seeds. 
(4)~Both \reviscaIV{state-of-the-art} tools scale sublinearly as thread count increases, \reviscaIV{wasting available thread-level parallelism in hardware}.
These observations expose a pressing need to have a specialized, high-performance, scalable, and low-cost \reviscaIV{algorithm/hardware} co-design that alleviates bottlenecks in both the seeding and alignment steps of sequence-to-graph mapping.}

To this end, \textbf{our goal} is to design high-performance, scalable, power- and area-efficient hardware accelerators that alleviate bottlenecks in both the seeding and alignment steps of sequence-to-graph mapping, with support for \reviscaII{both short (e.g., Illumina~\cite{illuminawebsite,iseqwebpage,miniseqwebpage,miseqwebpage,nextseqwebpage,novaseqwebpage}) and long (e.g., PacBio~\cite{pacbiowebsite,sequelwebpage}, ONT~\cite{ontwebsite,minionwebpage,gridionwebpage,promethionwebpage}) reads.}}
\reviscaII{Since sequence-to-sequence (S2S) mapping can be treated as a special case of sequence-to-graph mapping (S2G), we aim to design a \emph{\reviscaII{universal} accelerator} that is effective and efficient for \emph{both} problems (S2G and S2S mapping).} 

We propose \mech, 
\revISCA{a \emph{universal genomic mapping accelerator} that supports both \underline{se}quence-to-\underline{gra}ph \underline{m}apping and sequence-to-sequence mapping, for both short and long reads.} \reviscaIII{\mech consists of two main components: (1)~\ms, the \emph{first} \underline{min}imizer-based \underline{seed}ing accelerator, which finds the candidate \reviscaIV{mapping locations (i.e., subgraphs)} in a given genome graph; and (2)~\ba, the \emph{first} \underline{bit}vector-based sequence-to-graph \underline{align}ment accelerator, which performs alignment between a given read and the subgraph identified by \ms. \ms is built upon a memory-efficient minimizer-based seeding algorithm, and \ba is built upon our \emph{novel} bitvector-based, highly-parallel sequence-to-graph alignment algorithm. 
}

\damlaISCA{
In \ms, the minimizer-based \reviscaII{seeding} approach
decreases the memory footprint \reviscaII{of the index 
and provides speedup during
\reviscaIV{seed queries.}} 
\reviscaIII{
\ms logic requires \sgc{only} \gagan{basic operations} (e.g., comparisons, simple \damla{arithmetic} operations, scratchpad \reviscaIV{read-write operations) that are implemented with simple logic.} 
}
Due to frequent memory accesses required for fetching the 
seeds, \reviscaII{we couple} \ms with High-Bandwidth Memory (HBM)~\cite{hbm}
\reviscaIV{to enable low-latency and highly-parallel memory \reviscaIII{access}, which alleviates the memory \reviscaIV{latency} bottleneck.}

In \ba, \reviscaII{we design a} new bitvector-based \reviscaII{alignment approach, which is amenable \reviscaIV{to} efficient hardware acceleration.}
\reviscaIII{\ba employs a systolic-array-based \reviscaII{design} to circulate the internal data (i.e., bitvectors) generated by different processing elements, which provides scalability and \reviscaII{reduces \reviscaIV{both memory bandwidth and memory footprint}.}} In order to handle hops \reviscaII{(i.e., non-neighbor nodes in the graph-based reference)}, \ba provides a simple design that contains queue structures between each processing element, which store the most recently generated bitvectors. 
}

\vspace{2pt}
\textbf{Key Results.} 
\reviscaI{We compare \mech with \reviscaII{seven} state-of-the-art works: \reviscaII{S2G} mapping software (GraphAligner~\cite{rautiainen2020graphaligner} and vg~\cite{garrison2018variation}, which are CPU-based, and HGA~\cite{feng2021hga}, which is GPU-based), SIMD-based \reviscaII{S2G} alignment software (PaSGAL~\cite{jain2019accelerating}), and hardware accelerators for \reviscaII{S2S} alignment (the GACT accelerator in Darwin~\cite{turakhia2018darwin}, the SillaX accelerator in GenAx~\cite{fujiki2018genax}, and GenASM~\cite{cali2020genasm}). \reviscaIII{We find that:}
\reviscaII{(1)~\mech outperforms state-of-the-art S2G mapping tools by $5.9\times$/$3.9\times$ and $106\times$/$742\times$ for long and short reads, respectively, while reducing power consumption by $4.1\times$/$4.4\times$ and $3.0\times$/$3.2\times$. (2)~\ba outperforms \reviscaIV{the} \reviscaIII{state-of-the-art} S2G alignment tool by $41\times$--$539\times$ and \reviscaIII{three} S2S alignment \reviscaIII{hardware} accelerators by $1.2\times$--$4.8\times$. (3)~\ms can be employed for the seeding step of both S2G and S2S mapping pipelines.
}
}

\vspace{1pt}
This paper makes the following contributions:

\begin{itemize}[itemsep=0pt, topsep=0pt, parsep=0pt, leftmargin=*]
    \item \reviscaII{We introduce \mech, the first universal genomic mapping accelerator for both sequence-to-graph and sequence-to-sequence mapping. \mech is also the first} algorithm/hardware co-design for accelerating sequence-to-graph mapping. \mech
    alleviates performance bottlenecks
    \revISCA{of 
    graph-based genome sequence analysis.}
    \item We propose \ms, the first \nour{algorithm/hardware  co-design} 
    for minimizer-based seeding. \ms can be used for the seeding steps of \reviscaII{both S2G mapping and traditional S2S mapping.}
    \item We propose \ba, the first \nour{algorithm/hardware  co-design} 
    for sequence-to-graph alignment. \ba is based on a novel bitvector-based S2G alignment algorithm \sgh{that we develop}, and can be also used as a S2S aligner.
    \reviscaI{\item \mech provides \reviscaII{large ($1.2\times$--$742\times$) performance and power benefits over seven state-of-the-art works} for \reviscaIII{end-to-end S2G mapping,} multiple steps of the S2G mapping pipeline, as well as the traditional S2S mapping pipeline.}
    \reviscaII{\item To aid research and reproducibility, we open source our software implementations of the \mech algorithms \reviscaIII{and datasets}~\cite{segramgithub}.}
\end{itemize}

\section{Background} \label{sec:background}

We present a brief background on the genome sequence analysis pipeline, and the changes required to \reviscaII{it} to support genome graphs.

\vspace{-6pt}
\zulalI{\subsection{Genome Sequence Analysis} \label{sec:gsa}
\hspace{\parindent}\textbf{Read Mapping.} Most \reviscaII{types of} genome sequence analysis start with finding the original locations of the \reviscaII{sequenced} reads on the reference genome of the organism, \reviscaII{via} a computational process called \emph{read mapping}~\cite{li2018minimap2,li2013aligning,alkan2009personalized,Xin2013,xin2015shifted,langmead2012fast,li2009soap2,alser2020sneakysnake,alser2020accelerating,alser2021technology}. 
To complete this task accurately in the shortest amount of time, many existing read mappers adopt a \emph{seed-and-extend} \reviscaIII{approach} that consists of four stages \reviscaII{(See Figure~\ref{fig:mapping-pipeline})}: \emph{indexing}, \emph{seeding}, optional \emph{filtering/chaining/clustering}, and \emph{alignment}. \emph{Indexing}~\circlednumber{\circleblack{.}0\circleblack{.}} pre-processes the reference genome \reviscaII{and generates an index of the reference to be later used in the next steps of read mapping.} \emph{Seeding}~\circlednumber{\circleblack{.}1\circleblack{.}} finds the set of \reviscaII{\emph{k}-length substrings (i.e., \emph{k-mers)} to represent each read and finds the exact matching locations of these k-mers in the reference genome (i.e., \emph{seeds}). These seeds from the reference genome represent the candidate mapping locations of the \reviscaIII{query read in the reference genome.}} Many read mappers include \reviscaII{an optional \emph{filtering/chaining/clustering} step~\circlednumber{\circleblack{.}2\circleblack{.}}} to eliminate \reviscaIII{candidate mapping regions around the seed locations from the previous step that are dissimilar to the query read} to decrease the number of alignment operations. Finally, to find the read's optimal mapping location while taking sequencing errors and the differences caused by variations and mutations into \reviscaIII{account}, \emph{alignment}~\circlednumber{\circleblack{.}3\circleblack{.}} performs approximate string matching (i.e., \emph{ASM}) between the read and the \reviscaII{reference regions around the non-filtered candidate mapping locations from the previous step. As part of \reviscaIII{the alignment} step, traceback is also performed to find the \emph{optimal alignment} between the read and the reference region, which is the alignment \reviscaIII{with the highest likelihood of being correct} (based on a scoring function~\cite{gotoh1982improved,miller1988sequence,waterman1984efficient}) or \reviscaIV{with} lowest edit distance (i.e., total number of edits: substitutions, insertions, deletions)~\reviscaIV{\cite{levenshtein1966binary}}}.

\textbf{Approximate String Matching (ASM)} finds the similarities and differences \reviscaII{(i.e., substitutions, insertions, deletions)} between two strings~\cite{navarro2001guided,waterman1976some,myers1999fast,ukkonen1985algorithms}. Traditional \reviscaII{ASM} methods \reviscaII{use} dynamic programming (DP) based algorithms, such as Levenshtein distance~\cite{levenshtein1966binary}, Smith-Waterman~\cite{smith1981identification}, and Needleman-Wunsch~\cite{needleman1970general}. Since DP-based algorithms have quadratic time and space complexity (i.e., O(m × n) between two sequences with lengths m and n), there is a dire need \reviscaII{for lower complexity algorithms or \reviscaIV{algorithm/hardware} co-designed ASM accelerators.}
One lower-complexity approach to ASM is bitvector-based algorithms, such as Bitap~\cite{wu1992fast, baeza1992new, cali2020genasm} and the Myers' algorithm~\cite{myers1999fast}.}


\vspace{-6pt}
\subsection{Graph-Based Genome Sequence Analysis}

\hspace{\parindent}\textbf{Genome Graphs.}
Genetic variations between \reviscaII{two} individuals \reviscaII{are} observed by \sgh{comparing the differences between their two genomes.}
These differences, such as single-nucleotide polymorphisms (i.e., SNPs)~\cite{syvanen2001accessing,cargill1999characterization}, insertions and deletions (i.e., indels), and structural variations (i.e., SVs)~\reviscaIII{\cite{feuk2006structural,sudmant2015integrated,alkan2011genome,ho2020structural}}, \reviscaII{lead to} genetic diversity between populations and within communities~\cite{10002015global}. However, the presence of these genomic \reviscaII{variations} creates limitations when mapping the sequenced reads to a reference genome~\cite{paten2017genome, degner2009effect, brandt2015mapping, gunther2019presence}, since the reference \reviscaIII{genome} is commonly represented as a \reviscaIII{single \emph{linear}} \reviscaII{DNA sequence, which does not reflect all the genetic variations that exist in a population}~\cite{schneider2017evaluation}. \sgh{Using a} single reference \reviscaIII{genome} introduces \emph{reference bias}, \reviscaII{by \emph{only} emphasizing the genetic variations that are present in the \reviscaIII{single} reference \sgh{genome}~\cite{computational2018computational,paten2017genome,vernikos2015ten,sherman2020pan,liu2020one,golicz2020pangenomics} \reviscaIII{and ignoring other variations that are not represented in the single linear reference sequence}. 
These factors lead to low read mapping accuracy around the \reviscaIII{genomic regions that have SNPs, indels and SVs, and eventually cause, for example, false detection of SVs~\cite{rakocevic2019fast}.}} 

\reviscaII{Genome graphs} are better suited for expressing the \reviscaII{the \reviscaIII{genomic regions that have SNPs, indels and SVs} than a linear reference sequence~\cite{garrison2018variation} since genome graphs combine the linear reference genome with the \emph{known genetic variations} in the \nour{entire} population as a graph-based data structure.} Therefore, there is a \sgh{growing trend} towards \reviscaII{using} genome graphs~\cite{pevzner2001eulerian, nurk2022complete, rautiainen2020graphaligner, garrison2018variation, rakocevic2019fast, kim2019graph, dilthey2015improved, olson2021precisionfda,zhang2020comprehensive,li2020design} \reviscaII{to more accurately express the genetic diversity in a population.} 
\damlaISCA{With increasing importance and usage of genome graphs, having \reviscaII{accurate and} efficient tools for \damla{mapping genomic} sequences to these graphs \reviscaII{has} become crucial.}

\textbf{Sequence-to-Graph Mapping.}\label{sec:background-mapping}
Similar to \damlaISCA{traditional} sequence-to-sequence mapping (Section~\ref{sec:gsa}), sequence-to-graph mapping also follows the \emph{seed-and-extend strategy}. \reviscaII{Sequence-to-graph mapping pipeline has two pre-processing and three main steps \reviscaIII{(see Figure~\ref{fig:mapping-pipeline}). The first pre-processing step constructs the genome graph~\circlednumber{0.1} using a linear reference genome and the associated variations for that genome. The second pre-processing step indexes the nodes of the graph~\circlednumber{0.2}. The resulting index is used in the first main step of the pipeline, \emph{seeding}~\circlednumber{\circleblack{.}1\circleblack{.}}, which aims to find seed matches between the query read and a region of the graph. After optionally filtering these seed matches with a \emph{filtering}~\cite{xin2015shifted,alser2017gatekeeper,kim2018grim}, \emph{chaining}~\cite{li2016minimap,li2018minimap2,li2020design,kalikar2022accelerating}, or \emph{clustering}~\cite{garrison2018variation,rautiainen2020graphaligner} step~\circlednumber{\circleblack{.}2\circleblack{.}}, \emph{alignment}~\circlednumber{\circleblack{.}3\circleblack{.}} is performed between all of the non-filtered seed locations within} the graph and the query read.} Even though sequence-to-sequence mapping is a well-studied problem, given the additional complexities and overheads of processing a genome graph instead of a linear reference genome \reviscaII{(\reviscaIII{see} Section~\ref{sec:motivation}),} sequence-to-graph mapping is a more difficult computational problem with \reviscaII{a \reviscaIII{smaller} number of} practical \sgh{software} tools \sgh{currently} available.



\begin{figure}[h!]
\centering
\vspace{-8pt}
\includegraphics[width=\columnwidth,keepaspectratio]{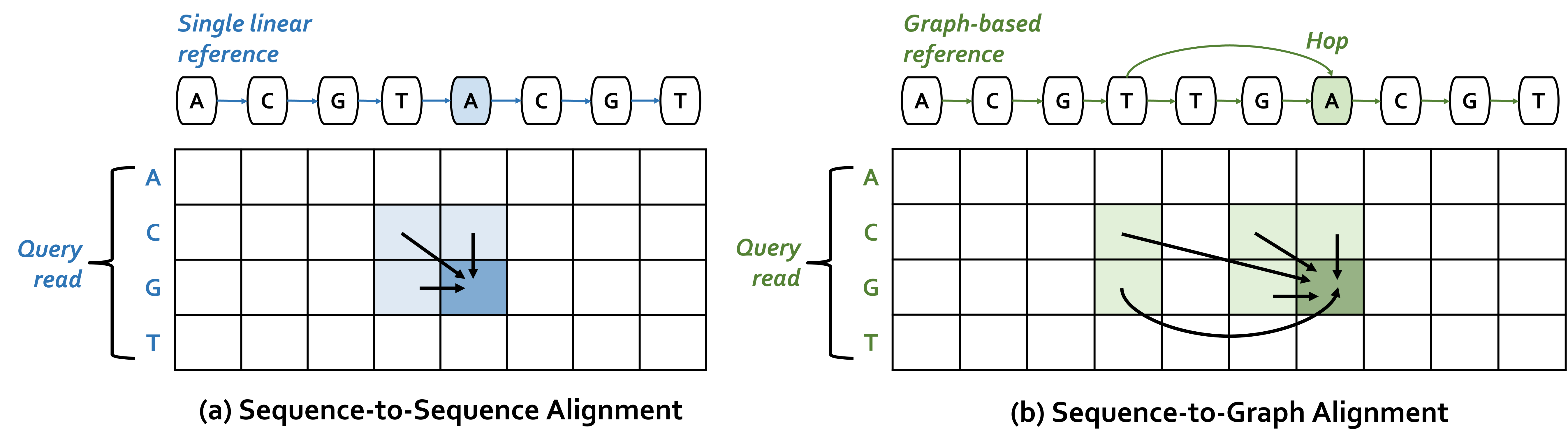}
\vspace{-20pt}
\caption{Data \reviscaIII{dependencies} in (a)~sequence-to-sequence alignment, and (b)~sequence-to-graph alignment.}
\label{fig:graph-table-dependencies}
\vspace{-8pt}
\end{figure}

\textbf{Sequence-to-Graph Alignment.}
The goal of aligning a sequence to a graph is \reviscaIII{to find} the path on the graph \reviscaIII{with the highest likelihood of being correct~\cite{jain2019accelerating}}. 
Similar to traditional \reviscaII{sequence-to-sequence (S2S) alignment, sequence-to-graph (S2G) alignment} also employs DP-based algorithms with quadratic time complexity~\cite{navarro2001guided,jain2020complexity,kavya2019sequence,gao2021abpoa,jain2019accelerating}. \reviscaIII{
\reviscaIV{A DP-based algorithm operates on a table,} where \reviscaIV{each column of the table corresponds to a reference character, and each row of the table corresponds to a query read character. Each} cell of the table can be interpreted as the cost of a partial alignment \damlaISCA{between the subsequences of the reference and \reviscaIV{ of the query read} that have been traversed so far}. \reviscaIV{In S2S alignment}, a new cell in the table is determined with simple rules from 3 of its neighbor cells. \reviscaIII{For example, as we show in Figure~\ref{fig:graph-table-dependencies}a, when computing the blue-shaded cell, we need information only from the three light blue-shaded cells.} In contrast to S2S alignment, S2G alignment must incorporate \sgc{non-neighboring} characters as well whenever there is an edge \reviscaIV{(i.e., \emph{hop})} from the non-neighboring character to the current character. For example, as we show in Figure~\ref{fig:graph-table-dependencies}b, when computing the green-shaded cell, we need \sgc{information from \emph{all} of the} light green-shaded cells. 
}

Even though there are many efforts for optimizing or accelerating the DP-based algorithms for S2S alignment~\cite{turakhia2018darwin,fujiki2018genax, fujiki2020seedex,haghi2021fpga,fei2018fpgasw,banerjee2018asap}, obtaining efficient solutions for S2G alignment demands attention with the growing \reviscaIV{usage of} genome graphs for genome sequence analysis.
\section{Motivation and Goal}
\label{sec:motivation}

\reviscaII{
\subsection{Software Tool Analysis}\label{sec:sw_tool_analysis}
In order to \reviscaIII{understand} the performance bottlenecks of the state-of-the-art sequence-to-graph mapping tools, we \reviscaIII{rigorously} analyze two such tools, GraphAligner~\cite{rautiainen2020graphaligner} and vg~\cite{garrison2018variation}, running on \reviscaIII{an} Intel\textsuperscript{\textregistered} Xeon\textsuperscript{\textregistered} E5-2630 v4 CPU~\cite{intel_cpu} \reviscaIII{with 20 physical cores/40 logical cores with hyper-threading~\reviscaIV{\cite{magro2002hyper,koufaty2003hyperthreading,tullsen1995simultaneous,tullsen1996exploiting}}}, operating at 2.20GHz, with 128GB DDR4 memory. Based on our bottleneck analysis with Intel VTune~\cite{intelvtune} \reviscaIII{and Linux Perf Tools~\cite{perftool}}, we make four key observations.}

\reviscaIII{\textbf{Observation 1: Alignment Step is the Bottleneck.}} Among the \reviscaIII{three main} steps of the sequence-to-graph mapping pipeline \reviscaIII{(Figure~\ref{fig:mapping-pipeline})}, 
the alignment step constitutes 50--95\% of the end-to-end execution \reviscaIII{time} \reviscaII{(measured across three short and four long read datasets; \reviscaIII{see Section~\ref{sec:gengraph-methodology:datasets}}). As shown in prior works~\cite{turakhia2018darwin, fujiki2018genax, cali2020genasm, fujiki2020seedex, alser2020accelerating},} sequence-to-sequence alignment is one of the major bottlenecks of the genome sequence analysis pipeline, and needs to be accelerated using specialized hardware. Since a graph-based representation of the genome is more complex than the linear representation, sequence-to-graph alignment places greater pressure on this bottleneck. 

\reviscaIII{\textbf{Observation 2: Alignment Suffers from High Cache Miss Rates.}} 
GraphAligner \revISCA{has a cache miss rate\reviscaIV{\footnote{We use the \emph{cache-misses} metric from Linux Perf Tools~\cite{perftool}.}} of 41\%, \sgis{meaning that} GraphAligner requires improvements to the on-chip caches} (e.g., lower access latency) in order to improve its performance. We find that the main reason of this high \revISCA{cache miss rate} is the high amount of \reviscaII{intermediate} data that is generated and reused as part of the alignment step \reviscaII{(for the dynamic programming table). vg tackles this issue by dividing the read into overlapping chunks, \reviscaIII{which reduces the size of the dynamic programming table, thus the size of the intermediate data}}. 

\reviscaIV{\textbf{Observation 3: Seeding Suffers from \reviscaV{the} DRAM Latency Bottleneck.}} \reviscaII{Our profiling of the seeding step of the pipeline finds} that 
seeding requires a \sgis{significant number} of random main memory accesses \reviscaII{while querying the index for the seed locations and suffers from the DRAM latency bottleneck.} 

\reviscaIV{\textbf{Observation 4: Baseline Tools Scale Sublinearly.}} When we perform a scalability analysis by running GraphAligner and vg with 5, 10, 20, and 40 threads, we observe that both \sgis{tools scale} sublinearly (i.e., their parallel efficiency does not exceed 0.4). 
\revISCA{When we focus on the change in the cache miss rate with the number of threads (t=10, 20, 40), we observe that (1)~from t=10 to t=20 to t=40, the cache miss rate increases from 25\% to 29\% to 41\%, and (2)~76\% of cache misses are associated with the alignment step of sequence-to-graph mapping when t=40. These \reviscaIII{results} suggest that when the number of threads reaches the number of logical cores in a CPU system, due to the large amount of intermediate data required to be accessed in the caches during the alignment step, two threads sharing the same physical core experience significant cache \reviscaII{and main memory} interference with each other and cannot fully take advantage of \reviscaIII{the full} \reviscaII{thread-level} parallelism \reviscaIII{available in hardware}}.

When we take all \sgis{four} observations into \reviscaIII{account}, we \reviscaII{find} that we need to have a \reviscaIII{specialized, balanced, and scalable} design for compute units, on-chip memory, and main memory accesses for both \sgis{the} seeding and alignment steps of sequence-to-graph mapping. 
\sgis{Unfortunately, these bottlenecks cannot be solved easily by software-only or hardware-only solutions. Thus,}
there is a pressing need to \sgis{co-design new algorithms with new hardware to enable} \reviscaII{high-performance,} efficient, scalable, and low-cost sequence-to-graph mapping.

\vspace{-4pt}
\subsection{Accelerating Sequence-to-Graph Mapping}\label{sec:accelerating_s2g}

\hspace{\parindent}
\textbf{Sequence-to-Sequence Accelerators.}
Even though there are several hardware accelerators designed \sgh{to alleviate bottlenecks in} several steps of \reviscaI{traditional} \reviscaII{sequence-to-sequence (S2S) mapping} (e.g., pre-alignment filtering~\reviscaIII{\cite{kim2018grim,xin2015shifted,alser2017gatekeeper,alser2017magnet,alser2019shouji,alser2020sneakysnake,singh2021fpga,mansouri2022genstore,nag2019gencache,subramaniyan2020accelerating,huangfu2020nest,huangfu2019medal,laguna2020seed,cong2018smem++}, sequence-to-sequence alignment~\cite{turakhia2018darwin,fujiki2018genax,cali2020genasm,fujiki2020seedex,banerjee2018asap,kim2020geniehd,haghi2021fpga,khatamifard2021genvom,fei2018fpgasw,rucci2018swifold}), none of these designs can be directly employed} for the sequence-to-graph (S2G) mapping problem. \reviscaII{This is because S2S mapping is a special case of S2G mapping, where all nodes have only one edge (Figure~\ref{fig:graph-table-dependencies}a).} \sgis{Existing} accelerators are limited to \sgis{\emph{only}} this special case, \sgis{and are} unsuitable for the \reviscaII{more general \reviscaII{S2G mapping} problem, where we also need to consider multiple edges \reviscaIV{(i.e., hops)} that a node can have (Figure~\ref{fig:graph-table-dependencies}b).} 

\reviscaI{S2G mapping is a more complex problem than S2S mapping since the graph structure is more complex than a linear sequence. \sgc{This \reviscaIII{additional complexity} results in four issues.} First, even though solutions for both problems follow the seed-and-extend approach, the already-expensive alignment step of S2S mapping is even more expensive in S2G mapping due to the hops in the graph \sgc{that must be} handled. Second, these hops add irregularity to the execution flow of alignment since they can \sgc{originate} from any vertex in the graph, \reviscaIII{leading to more data dependencies and irregular memory accesses}. Third, the heuristics used in S2S alignment \sgc{are often not} directly applicable to the S2G problem, \reviscaIII{as they assume a single linear reference sequence}. For example, chaining, \sgc{which is} used to combine different seed hits in long read mapping \reviscaIII{(assuming they are part of a linear sequence)}, cannot be used \reviscaII{directly for a genome graph} because there can be \emph{multiple} paths connecting two seeds together in the graph. \sgc{Fourth}, since the genome graph contains \reviscaII{both the linear reference sequence and the genetic variations}, the search space for the query reads is much larger \reviscaIII{in S2G mapping} than in S2S mapping. 

Existing S2S mapping accelerators can mimic the behavior of S2G mapping by taking \reviscaIV{\emph{all}} \reviscaII{paths \reviscaIII{that} exist in the genome graph} into \reviscaIII{account} and aligning the same read to \reviscaIII{each of} these paths \reviscaIV{\emph{one at a time}}. However, this would be \reviscaII{prohibitively} inefficient \reviscaIII{in terms of both computation and memory requirements (e.g., it would require an exorbitant memory footprint to store all possible graph paths as separate linear sequence strings).}
Thus,} with the growing importance and usage of genome graphs, it is crucial to have efficient designs \reviscaII{optimized} for sequence-to-graph mapping, which \reviscaII{can effectively} work with both short and long reads. 

\textbf{Graph Processing Accelerators.}
\reviscaI{Unlike typical graph traversal workloads~\reviscaII{\cite{page1999pagerank,moore1959shortest,dijkstra1959note,su2009survey}}, sequence-to-graph mapping involves \reviscaIII{high amounts of} both random memory accesses (due to the seeding step) and expensive computations (due to the alignment step). 
\sgc{Seeding enables the mapping algorithm to detect and focus on only certain candidate subgraphs, eliminating the need for a full graph traversal. Alignment is not a graph traversal workload, and instead is an expensive bitvector-based or DP-based \reviscaIII{computational problem}.}
\sgc{While existing} graph accelerators~\cite{zhong2013medusa,graphp,ahn2015pim,ahn2015scalable,graphpim,song2018graphr,sengupta2015graphreduce,nurvitadhi2014graphgen,gui2019survey,khoram2018accelerating,dai2018graphh,zhang2017boosting,zhang2015efficient,zhou2017tunao,ozdal2016energy,ham2016graphicionado,dai2017foregraph,zhou2016high,dai2016fpgp,attia2014cygraph,delorimier2006graphstep,wang2016gunrock,khorasani2014cusha,wang2019processor,besta2021sisa} could \sgc{potentially} \reviscaIII{be customized to} help the seeding step of the sequence-to-graph mapping pipeline, they are unable to handle the \reviscaIII{major bottleneck} of sequence-to-graph mapping, which is alignment.}

\vspace{-6pt}
\subsection{Our Goal} 
\reviscaIII{Our} goal is to design a high-performance, memory-efficient, and scalable hardware acceleration framework for sequence-to-graph mapping \reviscaII{that can also effectively perform sequence-to-sequence mapping}. To this end, we propose \mech, \reviscaI{the first \emph{universal genomic mapping accelerator} that can support both \underline{se}quence-to-\underline{gra}ph \underline{m}apping and sequence-to-sequence mapping, for both short and long reads.} To our knowledge, \mech is the first \reviscaIV{al\-go\-rithm/hard\-ware} co-design for accelerating sequence-to-graph mapping.
\vspace{-8pt}
\section{\mech: High-Level Overview} \label{sec:overview}

\begin{figure}[ht!]
\centering
\vspace{-6pt}
\includegraphics[width=\columnwidth,keepaspectratio]{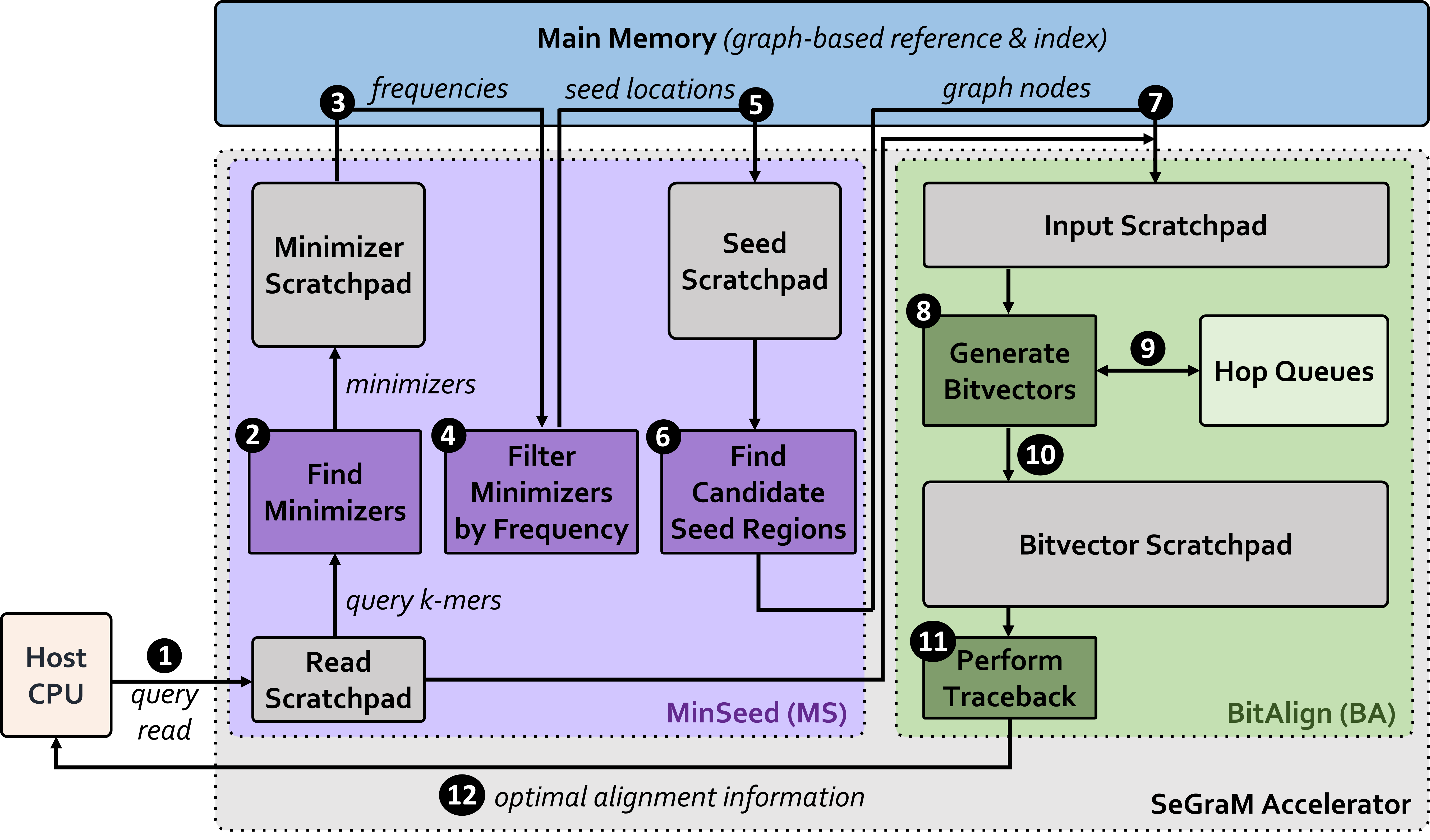}
\vspace{-18pt}
\caption{\reviscaII{Overview of \mech.}} 
\label{fig:gengraph-pipeline}
\vspace{-6pt}
\end{figure}

\reviscaII{\sgd{\mech provides efficient and general-purpose acceleration for} both the seeding and 
alignment steps of the sequence-to-graph mapping pipeline.
We base \mech upon a minimizer-based seeding algorithm and we propose a novel bitvector-based algorithm to perform approximate string matching between a read and a graph-based reference \reviscaII{genome}. We \emph{co-design} both algorithms with high-performance, scalable, and efficient hardware accelerators. As we show in Figure~\ref{fig:gengraph-pipeline}, \reviscaIV{a \mech accelerator} consists of two main components: \reviscaIV{(1)~\ms~(MS), which finds the minimizers for a given query read, fetches the candidate seed locations for the selected minimizers, and for each candidate seed, fetches the subgraph surrounding the seed; and (2)~\ba~(BA), which, aligns the query read to the subgraphs identified by \ms, and finds the optimal alignment.} To our knowledge, \ms is the first hardware accelerator for minimizer-based seeding and \ba is the first hardware accelerator for sequence-to-graph alignment.}

Before \mech execution starts, \sgd{pre-processing steps (1)~generate} each chromosome's graph structure, \sgd{(2)~index} each graph's nodes, and \sgd{(3)~pre-load} both the resulting graph and hash table index \sgd{into} the main memory. \reviscaII{Both the graph and its index are static data structures that can be generated \emph{only once} and reused for multiple mapping executions} \sgd{(Section~\ref{sec:preprocessing})}.

\mech execution starts when the query read is streamed from the host and \ms writes it to the \emph{read scratchpad} (\circlednumber{1}). \damlaISCA{Using all of the \emph{k}-length subsequences (i.e., \emph{k-mers}) of the query read, \ms finds the minimum representative set of these k-mers (i.e., \emph{minimizers})} according to a scoring mechanism and writes them to the \emph{minimizer scratchpad} (\circlednumber{2}). For each minimizer, \ms fetches its \sgd{occurrence} frequency from the hash table \sgd{in} main memory (\circlednumber{3}) and filters out \sgd{each minimizer whose occurrence} frequency is above a user-defined threshold (\circlednumber{4}). \reviscaII{We aim to select the least frequent \reviscaIV{minimizers} and filter out the most frequent minimizers such that we minimize the number of seed locations to be considered for the expensive alignment step.} Next, \ms fetches the seed locations of the remaining minimizers from main memory, and writes them to the \emph{seed scratchpad} (\circlednumber{5}). Finally, \ms calculates the candidate reference region \reviscaII{(i.e., subgraph surrounding the seed)} for each seed (\circlednumber{6}), fetches the graph nodes \sgd{from memory for each candidate region in the reference and writes the nodes} to the \reviscaII{\emph{input scratchpad}} \reviscaIII{of \ba.} (\circlednumber{7}). 
\sgd{\ba starts by reading the subgraph and the query read \reviscaIII{from the \emph{input scratchpad}}, and generates} the bitvectors (\circlednumber{8}) required for \reviscaII{performing approximate string matching and \reviscaIII{edit} distance calculation.} 
\reviscaIII{While generating these bitvectors, \ba writes them to the \emph{hop queues} (\circlednumber{9}) in order to handle the hops required for graph-based alignment, and also, to the \emph{bitvector scratchpad} (\circlednumber{10}) to be later used as part of the traceback operation}. Once \ba finishes generating and writing all the bitvectors, it starts reading them back from the \emph{bitvector scratchpad}, performs the traceback operation~(\circlednumber{11}), finds the optimal alignment between the subgraph and the query read, \reviscaII{and streams the optimal alignment information back to the host~(\circlednumber{12})}.
\section{Pre-Processing for \mech} \label{sec:preprocessing}

\sgc{\mech requires two pre-processing steps before it can start} execution: (1)~generating the graph-based reference, 
and (2)~\sgc{generating the hash-table-based index for the reference graph.} \reviscaII{After generating both data structures, we pre-load both the resulting graph and its index \sgd{into} main memory. Both the graph and its index are static data structures that can be generated \emph{only once} and reused for multiple mapping executions.} \reviscaIII{As such, pre-processing overheads are expected to be amortized across many mapping executions.}

\textbf{Graph-Based Reference Generation.}
As the first pre-pro\-cess\-ing step, we generate the graph-based reference using a linear reference genome (i.e., as a FASTA file~\cite{fasta}) and its associated variations (i.e., as \sgc{one or more VCF files}~\cite{vcf}). We use the vg toolkit's~\cite{garrison2018variation} \texttt{vg construct} command, \sgc{and} generate one graph for each chromosome.
For the alignment step of sequence-to-graph mapping, we need to make sure the nodes of \reviscaIII{each graph} are topologically sorted. Thus, 
we sort \reviscaIII{each graph} using the \texttt{vg ids -s} command. Then, we convert our VG-formatted graphs to GFA-formatted~\cite{gfa} graphs using the \texttt{vg view} command since GFA is easier to work with for the later steps of the pre-processing. 

\begin{figure}[h!]
\centering
\vspace{-4pt}
\includegraphics[width=\columnwidth,keepaspectratio]{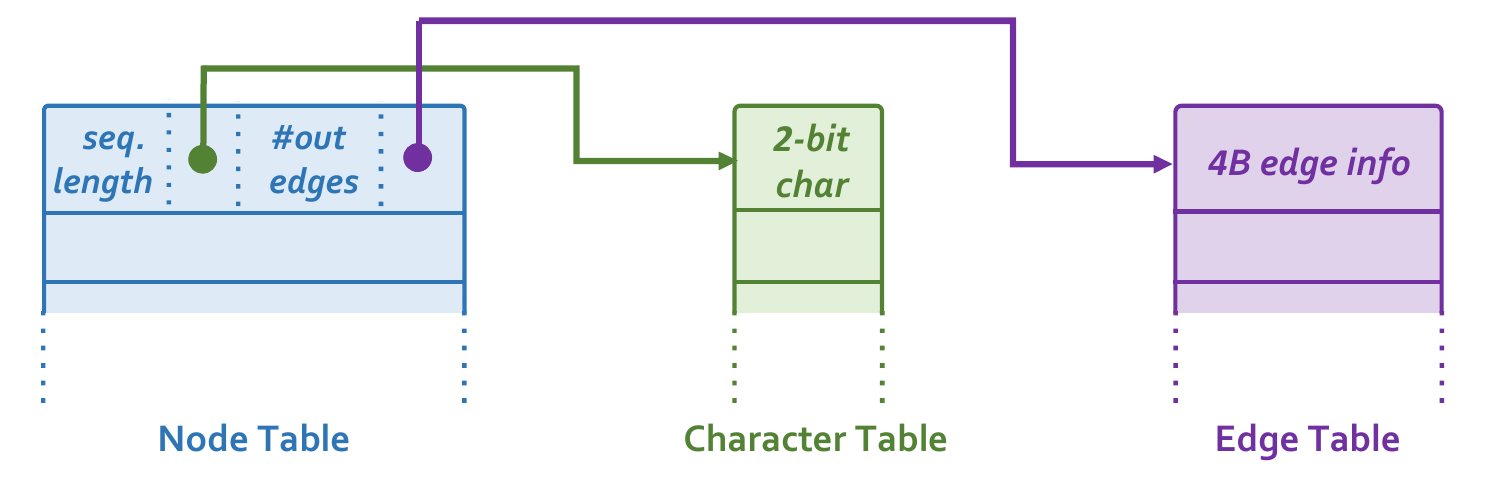}
\vspace{-20pt}
\caption{\reviscaII{Memory layout of the graph-based reference.} 
} 
\label{fig:gengraph-graphstructure}
\vspace{-4pt}
\end{figure}

As \sgd{shown} in Figure~\ref{fig:gengraph-graphstructure}, we generate three 
table structures to store the graph-based reference:
\sgc{(1)~the \emph{node table}, (2)~the \emph{character table}, and (3)~the \emph{edge table}.}
\sgc{The node table stores one entry for each \reviscaII{node of the graph}, using the node ID as the entry index, with the entry containing four fields: \reviscaIII{(i)~the length of the node sequence in characters, (ii)~the starting index corresponding to the node sequence in the character table, (iii)~the outgoing edge count for the node, and (iv)~the starting index corresponding to the node's list of outgoing edges in the edge table. The character table stores the associated sequence of each node, with each entry consisting of one character in the sequence (i.e., A, C, G, T).} \reviscaIV{The edge table stores the associated outgoing nodes of each node (\reviscaIII{indexed} by node ID), with each entry consisting of an outgoing node ID.}}

We use \sgc{statistics about each chromosome's associated graph} (i.e., number of nodes, number of edges, and total sequence length) 
to determine the \sgc{\sgd{size} of each table and of each table entry}.
Based on our analysis, we find that each \sgc{entry in the node table requires \SI{32}{\byte}, with a total table size of} \emph{\#nodes}~$*$~\SI{32}{\byte}. \damla{Since we can store characters in the \sgc{character} table using a 2-bit representation \reviscaIII{(A:00, C:01, G:10, T:11)}},
the total size \sgd{of the table} is \emph{total sequence length}~$*$~\SI{32}{bits}. \sgd{We find} that each entry in the \sgc{edge} table requires 4B, thus the total size of the edge table is \emph{\#edges}~$*$~\SI{4}{\byte}. \damlaI{\sgc{\sgd{Across} all 24 chromosomes \damlaISCA{(1--22, X, and Y)} of the human genome, the storage required for the \reviscaII{graph-based reference} is \SI{1.4}{\giga\byte}.}} \reviscaIII{We store the graph-based reference in main memory.}

\textbf{\sgc{Hash-Table-Based} Index Generation.}
As the second pre-pro\-cess\-ing step, we generate the \sgc{hash-table-based} index for each of the generated graphs \reviscaIII{(i.e., one index for each chromosome)}. 
The nodes of the graph structure are indexed and stored in the \reviscaIV{hash-table-based index}. As we explain in Section~\ref{sec:minseed_algo}, since \mech performs minimizer-based seeding, we use minimizers~\cite{roberts2004reducing,li2016minimap,li2018minimap2} as the \sgc{\reviscaII{hash table key}, and the minimizers' exact matching locations in the graphs' nodes as the \reviscaII{hash table value.}}

\begin{figure}[h!]
\centering
\vspace{-2pt}
\includegraphics[width=\columnwidth,keepaspectratio]{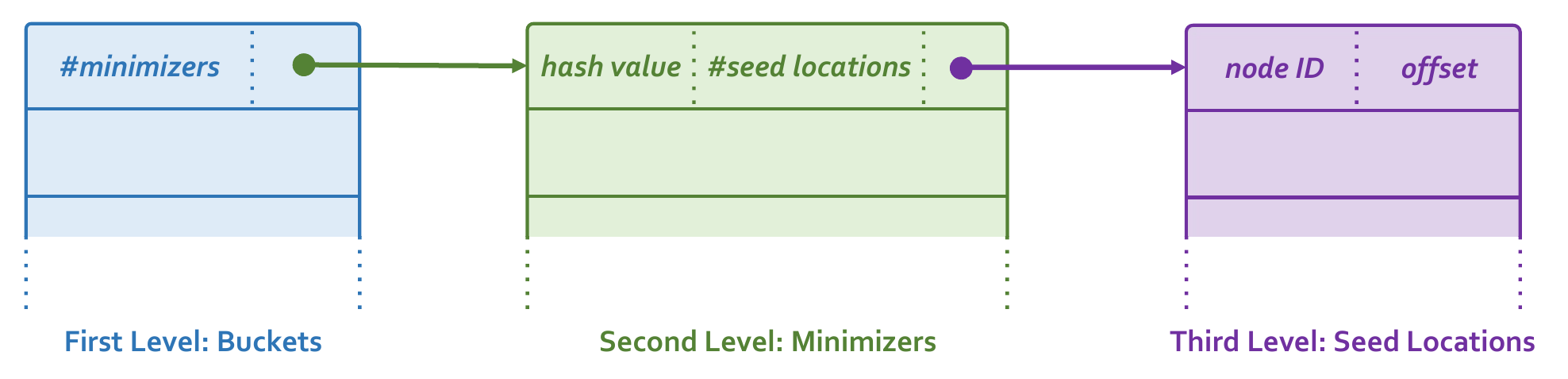}
\vspace{-22pt}
\caption{\reviscaII{Memory layout of the hash-table-based index.} 
} 
\label{fig:gengraph-indexstructure}
\vspace{-2pt}
\end{figure}

As \sgd{shown} in Figure~\ref{fig:gengraph-indexstructure}, we use a three-level structure to store the \sgc{hash-table-based} index. In the first level of the \reviscaIV{hash-table-based index}, similar to Minimap2~\cite{li2018minimap2}, we \sgc{use} \emph{buckets} to decrease the memory footprint of the index. Each entry in this \sgc{first level corresponds to a single bucket, and contains the starting address of the bucket's minimizers in the second-level table, along with the number of minimizers in the second-level table that belong to the bucket. In the second level, we store one entry for each \emph{minimizer}}. Each \sgc{second-level entry stores the hash value of the corresponding minimizer, the starting address of the minimizer's seed locations in the third-level table, and number of locations that belong to the minimizer. The minimizers} are sorted based on their hash values. \sgc{In the third level, each entry corresponds to one \emph{seed location}. An entry contains the node ID of the corresponding seed location, and the relative offset of the corresponding seed location within the node. Locations are grouped based on their corresponding minimizers,} and sorted within each group based on their values.

\begin{figure}[b!]
\centering
\vspace{-8pt}
\includegraphics[width=0.9\columnwidth,keepaspectratio]{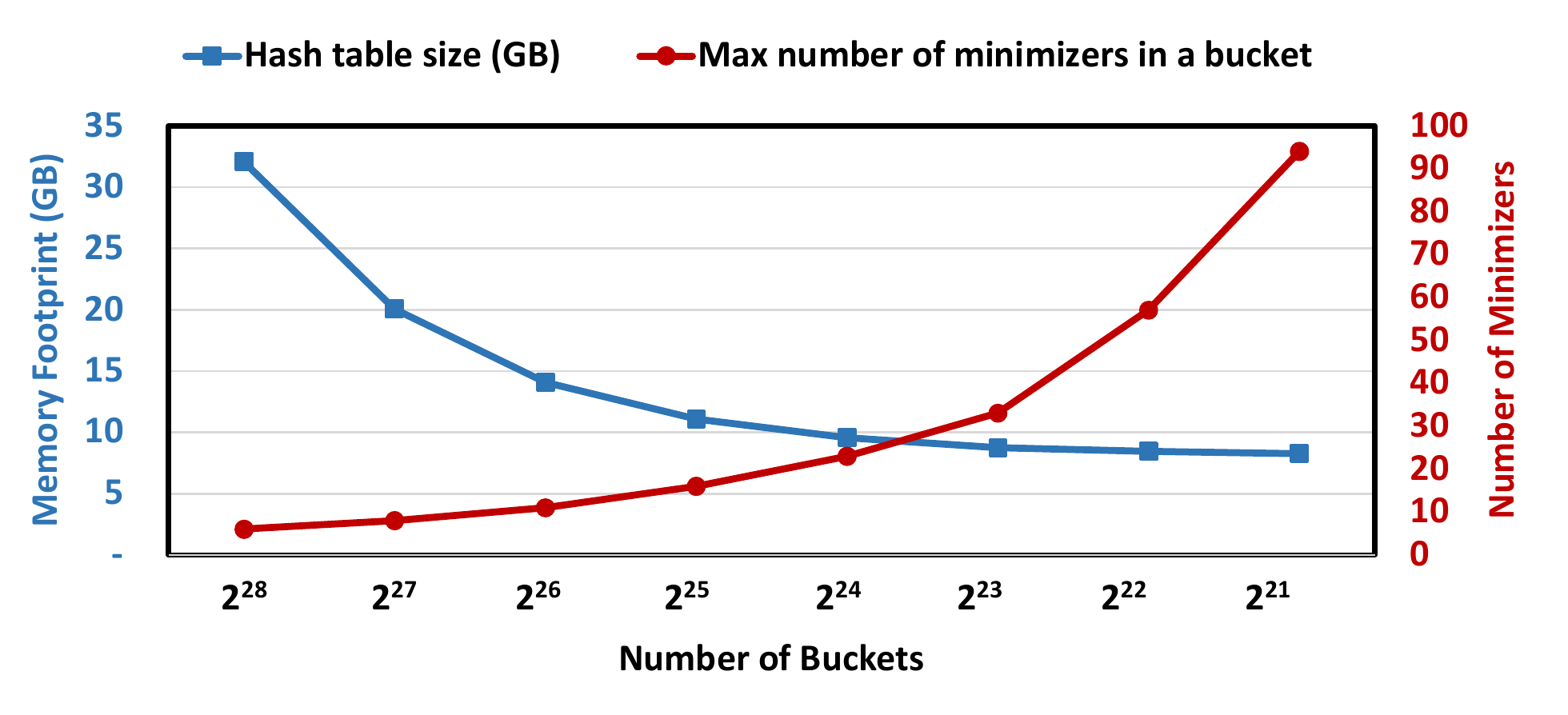}
\vspace{-12pt}
\caption{\reviscaIII{Effect of the bucket count on the memory footprint of the hash-table-based index and the \reviscaIII{maximum} number of minimizers per bucket.}
}
\label{fig:bucket-size}
\end{figure}

We use \sgc{statistics about each graph} (i.e., number of distinct minimizers, total number of locations, maximum number of minimizers per bucket, and maximum number of locations per minimizer) to determine the \sgc{size of the hash-table-based index}.
\sgc{We \sgd{empirically choose} the first-level bucket count.
\sgd{Figure~\ref{fig:bucket-size} shows the impact that the number of buckets has on both the \reviscaIII{total} memory footprint of the hash-table-based index (left axis, \reviscaIV{blue} squares) and the \reviscaIII{maximum} number of minimizers in each bucket (right axis, red dots). We observe from the figure that while}
a lower bucket count decreases the memory footprint of the index, \sgd{it increases} the number of minimizers assigned to each bucket (i.e., the number of hash collisions increases), increasing the number of memory lookups required.
We empirically \sgd{find} that a bucket count of $2^{24}$ strikes a reasonable balance.}
\sgc{Each} bucket entry requires \SI{4}{\byte} of data, \sgc{resulting in \sgd{a size} of $2^{24}$~$*$~\SI{4}{\byte} for the first level}. Each minimizer \sgc{requires \SI{12}{\byte} of data, resulting in \sgd{a size} of} \emph{\#distinct minimizers}~$*$~\SI{12}{\byte} \sgd{for the second level}. Each location \sgc{requires \SI{8}{\byte} of data, resulting in \sgd{a size} of} \emph{\#total number of locations}~$*$~\SI{8}{\byte} \sgd{for the third level}. \sgc{\sgd{Across} all 24 chromosomes \damlaISCA{(1--22, X, and Y)} of the human genome, the \sgd{total} storage required for the hash-table-based index is \SI{9.8}{\giga\byte}.} \reviscaIII{We store the hash-table-based index in main memory.}

\section{\ms Algorithm} \label{sec:minseed_algo}

\begin{figure}[ht!]
\centering
\vspace{-4pt}
\includegraphics[width=\columnwidth,keepaspectratio]{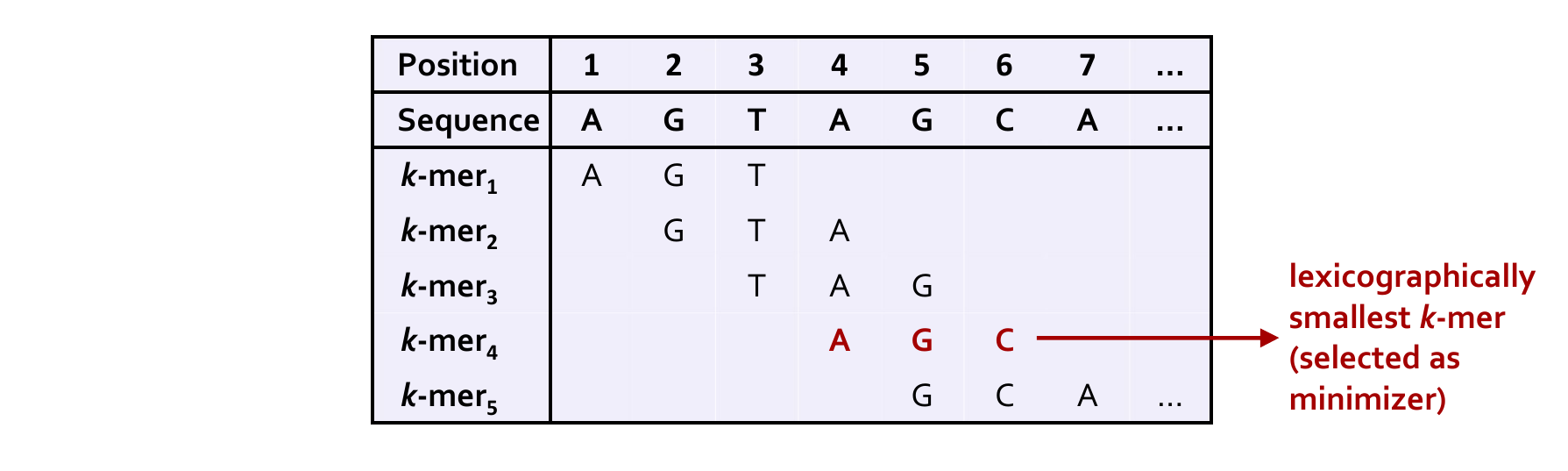}
\vspace{-16pt}
\caption{Example of finding \reviscaIII{the minimizer of the first window of a sequence.}} \label{fig:minimizers}
\vspace{-4pt}
\end{figure}

We base 
\sgc{\ms upon} Minimap2's minimizer-based seeding algorithm (i.e., \texttt{mm\_sketch}~\reviscaIII{\cite{li2016minimap,li2018minimap2,minimap2github}}). A \sgc{\emph{<w,k>}-minimizer}~\cite{roberts2004reducing,li2016minimap,li2018minimap2,schleimer2003winnowing,jain2020weighted,jain2022long} is the smallest \emph{k-mer} in a window of \emph{w} consecutive k-mers (according to a scoring mechanism), \sgc{for subsequences of length \emph{k}}. \damlaISCA{Minimizers ensure that two different sequences are represented with the same seed if they \reviscaII{share an exact match of at least $w+k-1$ bases long}}. \sgc{Compared to using the full set of k-mers, using only the \emph{<w,k>}-minimizers decreases the storage requirements of the index \reviscaII{(by a factor of $2/(w+1)$)} and speeds up index queries.}
In Figure~\ref{fig:minimizers}, we show an example of how \sgc{the} <5,3>-minimizer of a sequence is selected \sgc{from} the full set of k-mers \reviscaIII{from the sequence's first window}. After finding the 5 adjacent 3-mers, we sort them and select the smallest \reviscaII{based on a pre-defined ordering/sorting mechanism}. In this example, sorting is done \reviscaIII{based on lexicographical order and the lexicographically smallest k-mer is selected as the minimizer of the first window of the given sequence.}

\sgc{The} \ms algorithm starts \sgc{by} computing the minimizers of a given query read. 
\sgc{While a naive way to compute the minimizers is to use a nested loop (where the outer loop iterates over the query read to define each window \reviscaII{and} the inner loop finds the minimum k-mer \reviscaII{(i.e., minimizer)} within each window), we can eliminate the inner loop by caching the previous minimum k-mers \reviscaII{within the current window}. The single-loop algorithm has a complexity of $O(m)$, where $m$ is the length of the query read.}

After finding the minimizers \reviscaIII{of each read}, \ms queries the \sgc{hash-table-based index (Section~\ref{sec:preprocessing}) stored in} memory to fetch the \reviscaII{occurrence frequency} (i.e., \emph{\#locations}) of each minimizer. 
A minimizer is discarded if its \reviscaIII{occurrence frequency in the reference genome} is above a 
user-defined threshold 
\reviscaIV{(pre-computed for each chromosome in order to discard the top 0.02\% most \reviscaIII{frequent} minimizers), in order \reviscaII{to reduce the number of seed locations that are sent \reviscaIII{to the alignment step of the mapping pipeline~\cite{li2016minimap,li2018minimap2,Xin2013})}}}.
If the minimizer is not discarded, then all \sgc{of} the seed locations for that minimizer \sgc{are fetched from the index}.

\begin{figure}[ht!]
\centering
\vspace{-10pt}
\includegraphics[width=\columnwidth,keepaspectratio]{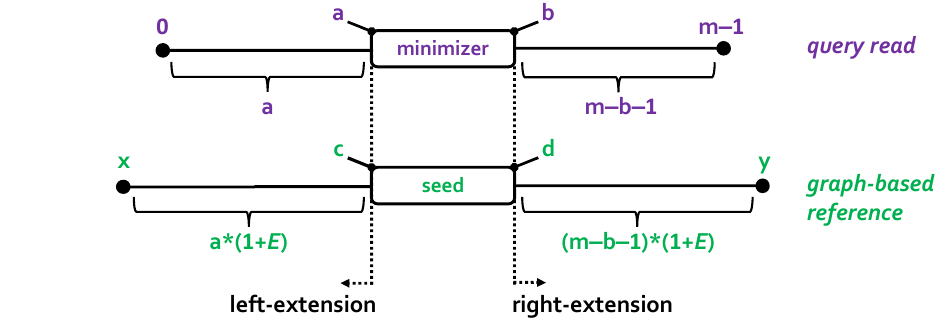}
\vspace{-20pt}
\caption{Calculations for finding the \reviscaIII{start ($x$) and end ($y$) positions of a candidate seed region (i.e., subgraph) using i)~the start ($a$) and end ($b$) positions of a minimizer within the query read and ii)~the start ($c$) and end ($d$) positions of a seed within the (graph-based) reference.}} \label{fig:seedregion}
\vspace{-6pt}
\end{figure}

After fetching all seed locations \reviscaIII{corresponding to all \reviscaIV{non-dis\-card\-ed} minimizers of a query read}, \sgc{\ms calculates the leftmost and rightmost positions of each seed, using the node ID and relative offset of the seed location along with the relative offset of the corresponding minimizer within the query read.}
\damla{As we show in Figure~\ref{fig:seedregion}, to find the leftmost position of the seed region ($x$), we need the start position of the minimizer within the query read ($a$), the start position of the seed \reviscaIII{within the (graph-based) reference} ($c$), and the error rate ($E$). Similarly, to find the rightmost position of the seed region ($y$), we need the end position of the minimizer within the query read ($b$), the end position of the seed \reviscaIII{within the (graph-based) reference} ($d$), the query read length ($m$), and the error rate ($E$).}
Finally, for all seeds of the query read, the subgraphs, which are found by using the calculated leftmost and rightmost positions of the seed regions, are fetched from \reviscaIV{main memory. These subgraphs serve} as the output of the \ms algorithm.

\vspace{-2pt}
\section{\ba Algorithm} \label{sec:bitgraph_algo}

\lijoel{After \ms determines \reviscaII{the subgraphs to perform alignment for each query read}, 
for each $(\texttt{read},$ $\texttt{subgraph})$ pair, \ba calculates the edit distance and the corresponding alignment between the two.} In order to provide an efficient, hardware-friendly, and low-cost solution, we \damla{\sgc{modify the sequence alignment algorithm \sgd{of} GenASM~\cite{cali2020genasm,genasmgithub}, which is bitvector-based, to support} sequence-to-graph alignment, and \sgc{we} exploit the \reviscaIII{\emph{bit-parallelism}} \sgc{that the GenASM} algorithm provides.}

\lijoel{\textbf{GenASM.}
GenASM~\cite{cali2020genasm} \sgd{makes} the bitvector-based Bitap algorithm~\cite{wu1992fast, baeza1992new} suitable for efficient hardware implementation. GenASM shares a common characteristic with the well-known DP-based algorithms~\reviscaIII{\cite{levenshtein1966binary,smith1981identification,needleman1970general}: both algorithms operate on tables (see Section~\ref{sec:background-mapping} and Figure~\ref{fig:graph-table-dependencies}a).} \damlaISCA{The key difference between GenASM-based alignment and DP-based alignment is that cell values are bitvectors in GenASM, whereas cell values are numerical values in DP-based algorithms.} In GenASM, the rules for computing new cell values \sgd{can be} formulated as \sgd{simple} bitwise operations, which are particularly \sgd{easy and} cheap to implement in hardware. \sgd{Unfortunately,} GenASM is limited to sequence-to-sequence alignment. \sgd{We build on GenASM's bitvector-based algorithm}
\damlaISCA{\sgc{to} develop our new sequence-to-graph alignment algorithm, \ba.}
}

\lijoel{\textbf{\ba.}}
\sgc{There is \sgd{a} major difference between se\-quence-to-se\-quence alignment and sequence-to-graph alignment:}  \sgd{for the current character, se\-quence-to-se\-quence alignment needs to know about only the neighboring (i.e., previous/adjacent) text character, whereas sequence-to-graph alignment must incorporate \sgc{non-neighboring} characters as well whenever there is an edge \reviscaIV{(i.e., hop)} \reviscaIII{from the current character to the non-neighboring character (see Section~\ref{sec:background-mapping} and Figure~\ref{fig:graph-table-dependencies}b)}}. 
To ensure \sgc{that each of these data dependencies can be resolved as quickly as possible}, \reviscaIII{we topologically sort the input graph 
during pre-processing, as described in Section~\ref{sec:preprocessing}.}

\sgd{Algorithm~\ref{bitalign-dc-alg} shows our new \ba algorithm}. 
\sgc{\ba starts \reviscaII{with a linearized and topologically sorted input subgraph.}
This ensures that \sgd{(1)~}we can iterate over each node of the input graph sequentially, and \sgd{(2)~}all of the required bitvectors for the current iteration have already been generated in previous iterations.} \reviscaII{Besides the subgraph, \ba also \reviscaIII{takes} the query read \reviscaIV{and the edit distance threshold (i.e., maximum number of edits to tolerate when performing approximate string matching~\cite{levenshtein1966binary}) as inputs.}}

{
\begin{algorithm}[ht!]
\footnotesize
\caption{BitAlign Algorithm}\label{bitalign-dc-alg}
\textbf{Inputs:} \texttt{\reviscaII{linearized and topologically sorted subgraph}} (reference), \\ \texttt{\reviscaII{query-read}} (pattern), \texttt{k} (edit distance threshold)\\
\textbf{Outputs:} \texttt{editDist} (minimum edit distance), \texttt{CIGARstr} (traceback output)
    \begin{algorithmic}[1]
        \State $\texttt{n} \gets \texttt{length of \revIII{linearized reference subgraph}}$
        \State $\texttt{m} \gets \texttt{length of \revIII{query \reviscaV{read}}}$
        \State $\texttt{PM} \gets $\texttt{genPatternBitmasks(\reviscaV{query-read})}
    \Comment{\comm{pre-process the \comm{query read}}}
        \State
        \State $\texttt{allR[n][d]} \gets \texttt{111...111}$
        \Comment{\comm{init R[d] bitvectors for all characters with 1s}}
        \State
        \For{\texttt{i in (n-1):\revV{-1:}0}}
    \Comment{\comm{iterate over each \comm{subgraph} node}}
        \State $\texttt{curChar} \gets \texttt{\reviscaV{subgraph}-nodes[i].char}$
        \State $\texttt{curPM} \gets \texttt{PM[curChar]}$
        \Comment{\comm{retrieve the pattern bitmask}}
        \State
        
        \State $\texttt{R0} \gets \texttt{111...111}$
        \Comment{\comm{status bitvector for exact match}}
        \For{\texttt{j in \reviscaV{subgraph}-nodes[i].successors}}
            \State $\texttt{R0} \gets \texttt{((R[j][0]}\verb|<<|\texttt{1) | curPM) \& R0}$ 
            \Comment{\comm{exact match calculation}}
        \EndFor
        \State $\texttt{allR[i][0]} \gets \texttt{R0}$
        \State
        
        \For{\texttt{d in 1:k }}
            \State $\texttt{I} \gets \texttt{(allR[i][d-1]}\verb|<<|\texttt{1)}$
            \Comment{\comm{insertion}}
            \State $\texttt{Rd} \gets \texttt{I}$
            \Comment{\comm{status bitvector for $d$ errors}}
            \For{\texttt{j in \reviscaV{subgraph}-nodes[i].successors}}
                \State $\texttt{D} \gets \texttt{allR[j][d-1]}$
                \Comment{\comm{deletion}}
                \State $\texttt{S} \gets \texttt{allR[j][d-1]}\verb|<<|\texttt{1}$
                \Comment{\comm{substitution}}
                \State $\texttt{M} \gets \texttt{(allR[j][d]}\verb|<<|\texttt{1)}\texttt{ | curPM}$
                \Comment{\comm{match}}
                \State $\texttt{Rd} \gets \texttt{D \& S \& M \& Rd}$
            \EndFor
            \State $\texttt{allR[i][d]} \gets \texttt{Rd}$
        \EndFor
    \EndFor
    \State $\texttt{<editDist, \reviscaIV{CIGARstr}>} \gets \texttt{traceback(allR, subgraph, \reviscaII{query-read})}$
    \end{algorithmic}
\end{algorithm}
}

\damlaISCA{Similar to GenASM, as a pre-processing step, we generate four pattern bitmasks for the query read (one for each character in the alphabet: A, C, G, T; Line~3). \sgc{Unlike in} GenASM, \sg{which stores only} the most recent status bitvectors (i.e., \emph{R[d] bitvectors}, where $0 \leq d \leq k$) that hold the partial match information, \sgc{\ba needs to store \emph{all} of the status bitvectors for \emph{all} of} the text iterations \reviscaIV{(i.e., \emph{allR[n][d]}, where $n$ is the length of the linearized reference subgraph; Line~5). These allR[n][d]} bitvectors will be later used by the traceback step of \ba (Line~25).

\sgc{Next, \ba iterates} over each node of the linearized graph (Line~7) and \sgc{retrieves} the pattern bitmask for each \reviscaII{node}, based on the character stored in the current node (Lines~8--9). \sgc{Unlike in GenASM,} when computing \reviscaII{three of the intermediate bitvectors (i.e., match, substitution, and deletion; \reviscaIII{Lines~11--14, 19--22)}}, \sgc{\ba incorporates} the hops as well by \reviscaIII{examining} all successor nodes that the current node has (Lines~12 and 19). When calculating the deletion ($D$), substitution ($S$), and match ($M$) bitvectors, we take the hops into consideration, whereas when calculating the insertion ($I$) bitvector \reviscaII{(Lines~17--18)}, we do \emph{not} need to, \reviscaIII{since an insertion does \emph{not} consume a character from the reference subgraph, but does so from the query read \emph{only}}. 
After computing all \sgc{of} these intermediate bitvectors, we store \sgc{only} the R[d] bitvector \reviscaIV{(i.e., ANDed version of the intermediate bitvectors)} \reviscaIII{in memory} (Lines~23--24). After completing \sgc{all iterations}, we perform traceback (Line~25) by traversing the stored bitvectors in the opposite direction \reviscaIII{to find the optimal alignment (based on the \reviscaIV{user-supplied} alignment scoring function).}}

\sgc{To perform GenASM-like traceback, \ba stores $3(k+1)$ \sgd{bitvectors per} graph edge (similar to how GenASM stores three out of the four intermediate vectors), \reviscaIV{where $k$ is the edit distance threshold}.}
Since the number of edges in the graph can only be bounded very loosely, the potential memory footprint \damla{increases} significantly, which \damla{is} expensive to implement in hardware. We solve this problem by storing only $k+1$ bitvectors per \emph{node} \damlaISCA{(i.e., R[d] bitvectors),} from which the $3(k+1)$ bitvectors per \emph{edge} can be regenerated on-demand during traceback. While this \reviscaIII{modification} incurs \reviscaIII{small additional computational overhead, it decreases} the memory footprint of the algorithm by \reviscaII{at least} $3\times$.
Since \sgc{the main area and power cost of the alignment hardware comes from \sgd{memory, we find this trade-off} favorable}.

\reviscaII{Similar to GenASM, \ba also follows the divide-and-conquer approach, where we divide the linearized subgraph and the query read into overlapping windows and execute \ba for each window. After all windows' traceback outputs are found, we merge them to find the final traceback output. This approach helps us to decrease the memory footprint and lower the complexity of the algorithm.}

\section{\mech Hardware Design} \label{sec:bitgraph_hw}

In \mech, we \emph{co-design} our new \ms and \sgc{\ba algorithms}
with specialized custom accelerators \reviscaIII{that can support both sequence-to-graph and sequence-to-sequence alignment}.

\subsection{\ms Accelerator}
\label{sec:minseed_hw}

\reviscaIII{As shown \reviscaIV{earlier} in Figure~\ref{fig:gengraph-pipeline}, the \ms accelerator consists of: (1)~three computation \sgd{blocks} responsible for finding the minimizers from a query read \reviscaIII{(\circlednumber{2} from Figure~\ref{fig:gengraph-pipeline})}, filtering \sgd{minimizers based on their frequencies} \reviscaIII{(\circlednumber{4})}, and finding the \reviscaII{candidate reference regions (i.e., subgraphs surrounding the candidate seeds) \reviscaIII{(\circlednumber{6})}}} 
by calculating \sgc{each seed's leftmost and rightmost} positions \reviscaII{(Section~\ref{sec:minseed_algo})}; \reviscaIV{(2)~three scratchpad (on-chip SRAM memory) blocks};
and (3)~the memory interface, which handles the \reviscaIII{lookups of minimizer} frequency, seed location, and subgraph \sgd{from the attached \reviscaIII{main} memory stack \reviscaIII{(\circlednumber{3}, \circlednumber{5}, and \circlednumber{7} from Figure~\ref{fig:gengraph-pipeline})}}.

\begin{figure}[h!]
\centering
\vspace{-12pt}
\includegraphics[width=\columnwidth,keepaspectratio]{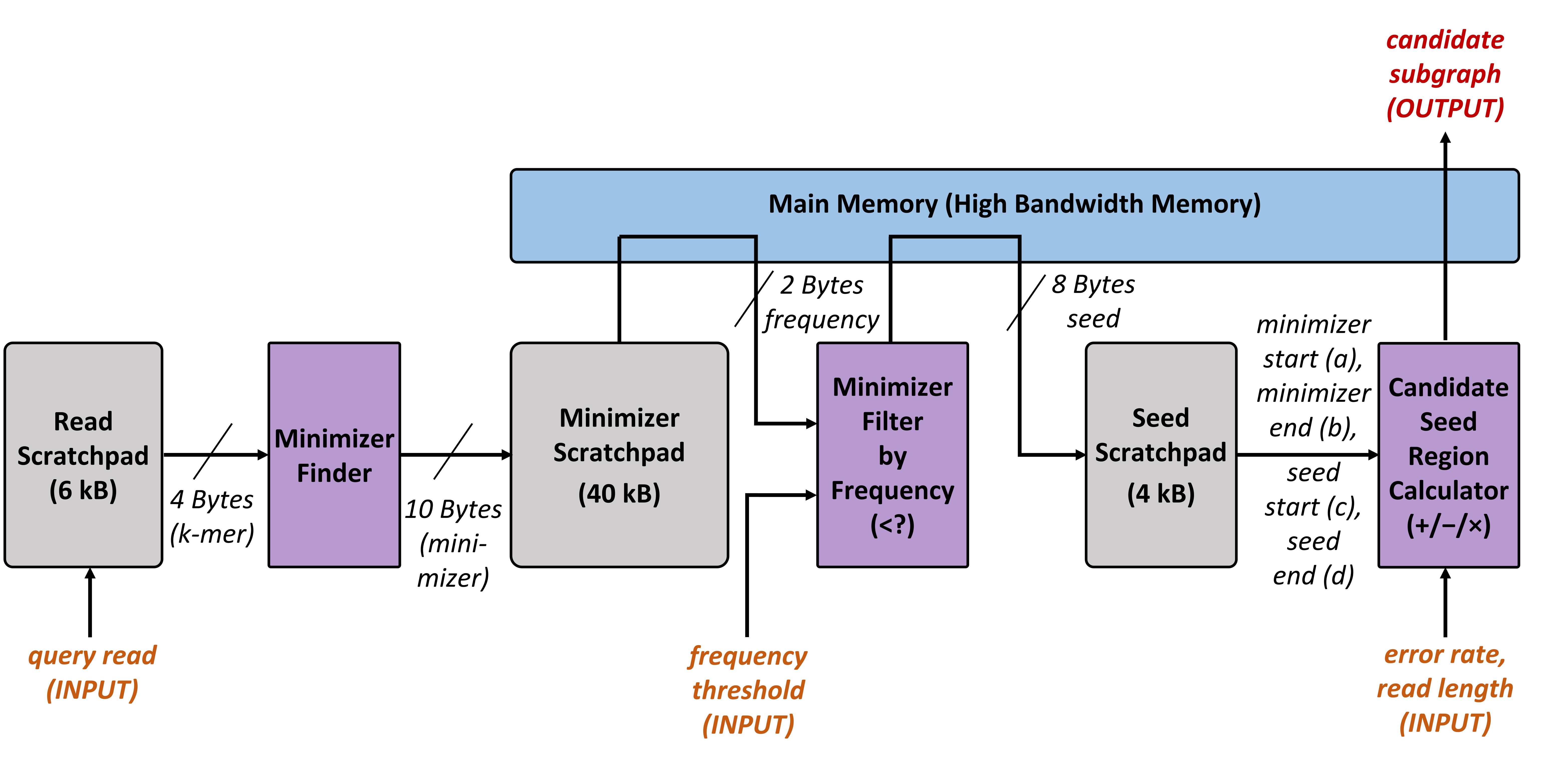}
\vspace{-20pt}
\caption{\sgd{\ms accelerator} design.}
\label{fig:minseed-hw}
\vspace{-6pt}
\end{figure}

As Figure~\ref{fig:minseed-hw} \sgd{shows}, \sgc{the} \ms accelerator \reviscaIII{receives} the query read as \sgc{its input,} and finds the candidate subgraphs \reviscaIII{(from the reference)} 
\reviscaII{as its} output. The computation modules \reviscaIII{(purple-shaded blocks)} are implemented with simple logic, since we require \sgc{only} \gagan{basic operations} (e.g., comparisons, simple \damla{arithmetic} operations, scratchpad R/W operations). 

\reviscaII{
\ms accelerator consists of three \reviscaIII{scratchpads} \reviscaIII{(gray-shaded blocks)}: (1)~\emph{read scratchpad}, which stores the query read; (2)~\emph{minimizer scratchpad}, which stores the minimizers fetched from the query read; and (3)~\emph{seed scratchpad}, which stores the seeds fetched from the memory for the minimizers. For all three scratchpads, we employ a double buffering technique to hide the latency of the \ms accelerator (See Section~\ref{sec:overall_hw}). Based on our empirical analysis, we find that (1)~the \emph{read scratchpad} requires \reviscaIV{\SI{6}{\kilo\byte}} of storage such that it can store 2 query reads of \reviscaIV{\SI{10}{\kilo\basepair} length}, \reviscaIV{where each character of the query reads is represented using 2 bits (A:00, C:01, G:10, T:11)}; (2)~the \emph{minimizer scratchpad} requires \reviscaIV{\SI{40}{\kilo\byte}} of storage such that it can store the minimizers of 2 different query reads, where the maximum number of minimizers that a query read in our datasets (Section~\ref{sec:gengraph-methodology:datasets}) can have is 2050 and each minimizer is represented with 10B of data; and (3)~the \emph{seed scratchpad} requires \reviscaIV{\SI{4}{\kilo\byte}} of storage such that it can store the seed locations of 2 different minimizers, where the maximum number of seed locations that a minimizer of a query read in our datasets can have is 242 and each seed location is represented with 8B of data.
}

\subsection{\ba Accelerator}
\label{sec:bitalign_hw}

\reviscaIII{As shown \reviscaIV{earlier} in Figure~\ref{fig:gengraph-pipeline}, the \ba accelerator} consists of: (1)~two computation blocks responsible for generating bitvectors to perform approximate string matching and \reviscaIII{edit} distance calculation \reviscaIII{(\circlednumber{8} from Figure~\ref{fig:gengraph-pipeline})} and traversing bitvectors to perform traceback \reviscaIII{(\circlednumber{11})}; (2)~two scratchpad blocks for storing the input data and intermediate bitvectors, and (3)~hop queue registers for handling the hops.

\begin{figure}[h!]
\centering
\vspace{-4pt}
\includegraphics[width=\columnwidth,keepaspectratio]{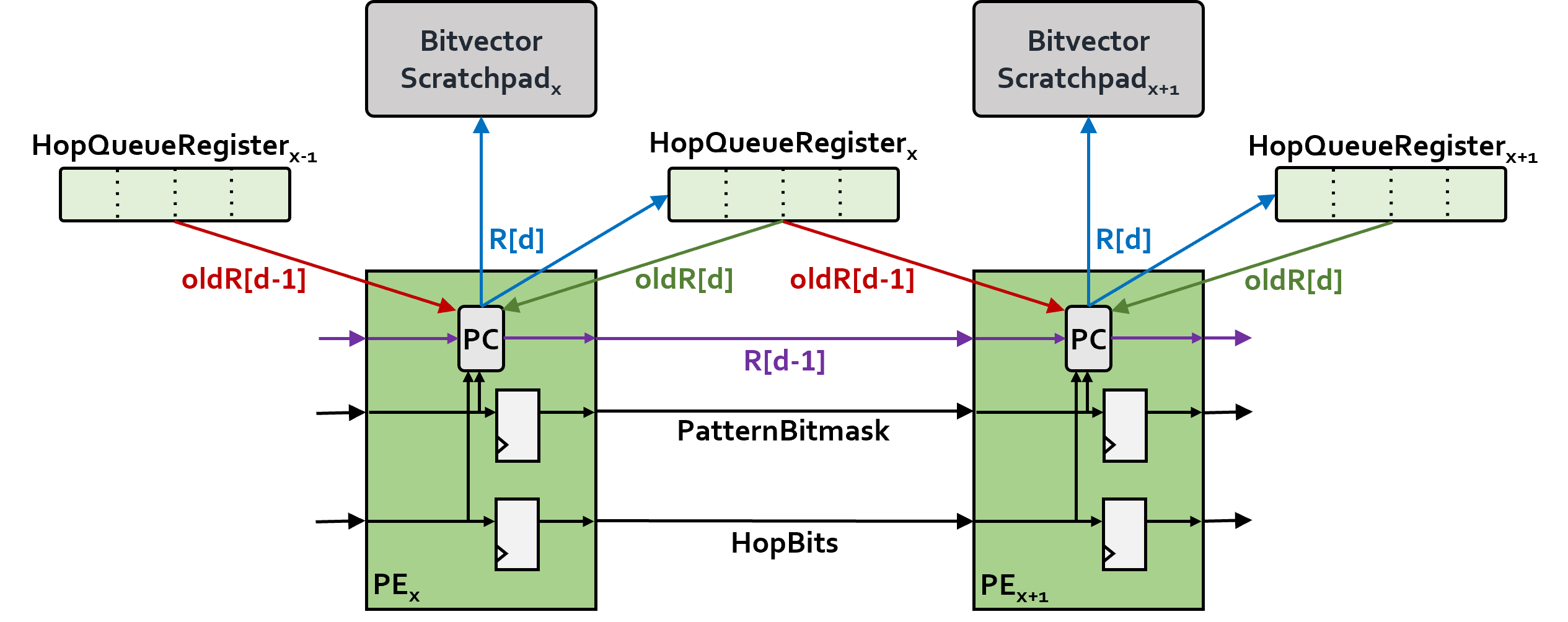}
\vspace{-14pt}
\caption{\sgc{Design of a \ba processing} element (PE).}
\label{fig:bitalign-hw}
\vspace{-4pt}
\end{figure}

We implement the \reviscaII{bitvector generation hardware} of \ba as a linear cyclic systolic array based~\cite{kung1978systolic,kung1982systolic} accelerator. 
\sgc{While this design is based \reviscaIII{on} the GenASM-DC hardware~\cite{cali2020genasm}, our new design incorporates}
\emph{hop queue registers} in order to feed the bitvectors of \sgc{non-neighboring} characters/nodes. As we show in Figure~\ref{fig:bitalign-hw}, the $R[d]$ bitvector \sgc{generated by} each processing element (PE) \reviscaII{(i.e., \emph{hop bitvector})} is fed to the tail of the hop queue register of the current PE. Each hop queue register then provides \sgc{its} stored bitvectors as the $oldR[d]$ bitvectors to the same PE (required for the match bitvector calculation), and as the $oldR[d-1]$ bitvectors to the next PE (required for deletion and substitution bitvector calculation) in the next cycle. \sgc{The} $R[d-1]$ bitvector (required for insertion bitvector calculation) is directly provided by the previous PE (i.e., not through the hop queue registers).

\begin{figure}[t!]
\centering
\includegraphics[width=0.8\columnwidth,keepaspectratio]{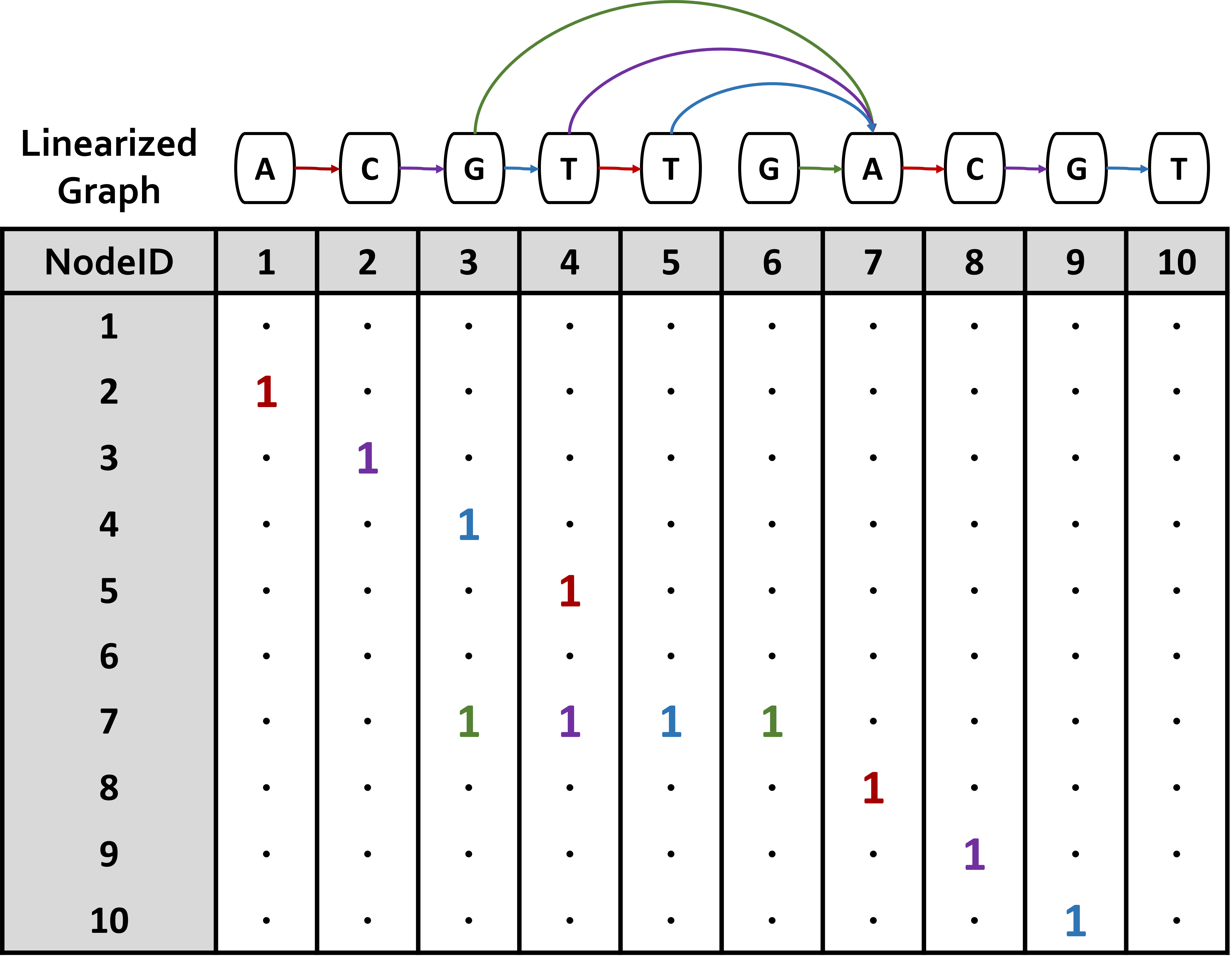}
\vspace{-10pt}
\caption{Linearized input subgraph and the 
HopBits \sgc{matrix}. \reviscaII{When there is an outgoing edge (hop) from $NodeID_x$ to $NodeID_y$, then the HopBits for the $y^{th}$ entry of $x^{th}$ column is 1, otherwise 0 \reviscaIII{(indicated as $\cdot$ for readability)}.}} 
\label{fig:bitalign-hopbits}
\vspace{-10pt}
\end{figure}

We implement the \reviscaII{hop information between nodes of the graph} as an adjacency matrix called \emph{HopBits} (Figure~\ref{fig:bitalign-hopbits}). 
Based on the HopBits \sgc{entry} of the current text character, either the actual hop bitvector \reviscaII{(if the HopBits entry is 1)}, \sgc{or a bitvector containing all ones \reviscaII{such that it will not have any effect on the bitwise operations (if the HopBits entry is 0)},} is used when calculating the match, deletion, and substitution bitvectors of the current PE. 

\begin{figure}[h!]
\centering
\vspace{-6pt}
\includegraphics[width=0.9\columnwidth,keepaspectratio]{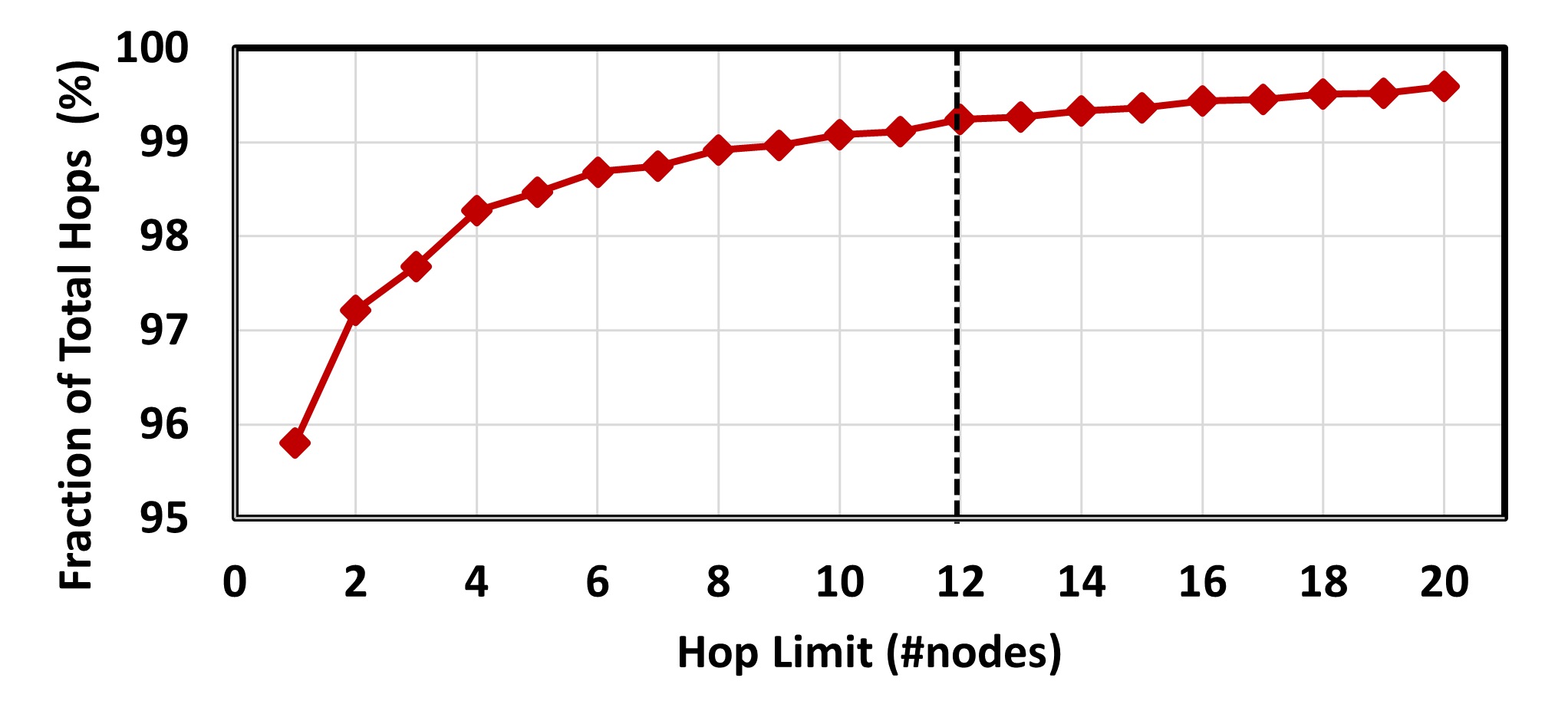}
\vspace{-12pt}
\caption{\sgc{Effect} of the hop limit on the fraction of hops included \reviscaIII{when performing sequence-to-graph alignment.}} \label{fig:bitalign-hoplimit}
\vspace{-6pt}
\end{figure}

In order to decrease the size of each hop queue register and the HopBits \sgc{matrix}, \reviscaIII{we perform an empirical analysis \reviscaIV{in Figure~\ref{fig:bitalign-hoplimit}, where we measure the fraction of total number of hops in the graph-based reference genome that we can cover} (Y-axis) when we limit the distance between the farthest node to take into account and the current node (i.e., \emph{hop limit}; X-axis). As Figure~\ref{fig:bitalign-hoplimit} shows, when we select 12 as the hop limit, \reviscaIV{we cover more than 99\% of all hops in the graph-based reference.} 
}
\reviscaI{\sgc{This is a reasonable limit because} a significant percentage of genetic variations in a human genome are either single nucleotide substitutions (i.e., single nucleotide polymorphisms, SNPs) or small indels (insertions/deletions)~\cite{10002015global}, which would result in short \reviscaII{hops} \sgc{that connect near-vicinity} vertices in the graph. On the other hand, even though large structural variations (SVs) would result in long \reviscaII{hops} connecting farther vertices in the graph, \sgc{such} SVs occur at \sgc{a very low} frequency.\reviscaIII{\footnote{Hop limit introduces a tradeoff between power/area overhead and accuracy. We leave overcoming this tradeoff and improving accuracy of our design as future work.}} As a result, in a topologically-sorted \reviscaIV{graph-based reference}, the majority of vertices are expected to have all \sgc{of} their neighbors within close proximity.}

\ba uses two types of SRAM buffers: \reviscaIV{(1)~\emph{input scratchpad}}, which stores the linearized reference graph, associated HopBits for each node, and the pattern bitmasks for the query read; \reviscaIV{and (2)~\emph{bitvector scratchpad}}, which stores the intermediate bitvectors generated during the \reviscaIII{edit distance calculation step of \ba and used during the traceback step of \ba.} For a \revISCA{64}-PE configuration with 128~bits of processing per PE \reviscaIII{(i.e., the width of bitvectors)}, \ba requires a total of \SI{24}{\kilo\byte} in \reviscaIII{\emph{input scratchpad} storage}. \sgc{Each PE also} requires a total of \SI{2}{\kilo\byte} \reviscaIII{\emph{bitvector scratchpad} storage (\SI{128}{\kilo\byte}, in total) and \SI{192}{\byte} \emph{hop queue register} storage (\SI{12}{\kilo\byte}, in total). In each cycle, \sgc{128~bits} of data (\SI{16}{\byte}) is written to each \reviscaIII{\emph{bitvector scratchpad}} and to each \emph{hop queue register} by each PE.}

\damla{As we explain in Section~\ref{sec:bitgraph_algo}, in order to decrease the memory footprint of the stored bitvectors required for traceback execution, we store \sgc{only} the ANDed version of the intermediate bitvectors ($R[d]$) and re-generate the intermediate bitvectors (i.e., match, substitution, deletion, and insertion) during \reviscaIII{traceback}. Thus, each element of the \reviscaIII{\emph{hop queue register}} has a length equal to the window size ($W$), instead of $3*W$. \reviscaIII{Similarly, with this design choice, the size of each \emph{bitvector scratchpad} \reviscaIV{of} each PE decreases by $3\times$}. 
}

\subsection{Overall System Design} \label{sec:overall_hw}


Figure~\ref{fig:overall-system} shows the overall design of \mech. 
\mech is connected to a host system
\reviscaI{\sgc{that} is responsible for the pre-processing steps \sgc{(Section~\ref{sec:preprocessing}) and for} transferring the query read to the accelerator. \reviscaIII{For fair comparison,} we exclude the time spent for generating and loading these structures to main memory for both our \mech design and for the baseline tools, as this is a one-time cost.
Since \sgd{pre-processing is} performed only once for each reference input, \sgd{it is not the bottleneck} of the \reviscaII{end-to-end execution}.\footnote{\reviscaIII{The reference input changes very infrequently and is reused for all query reads over likely millions of mapping invocations of \mech, allowing for the pre-processing latency to be amortized.}}

\begin{figure}[h!]
\centering
\vspace{-4pt}
\includegraphics[width=\columnwidth,keepaspectratio]{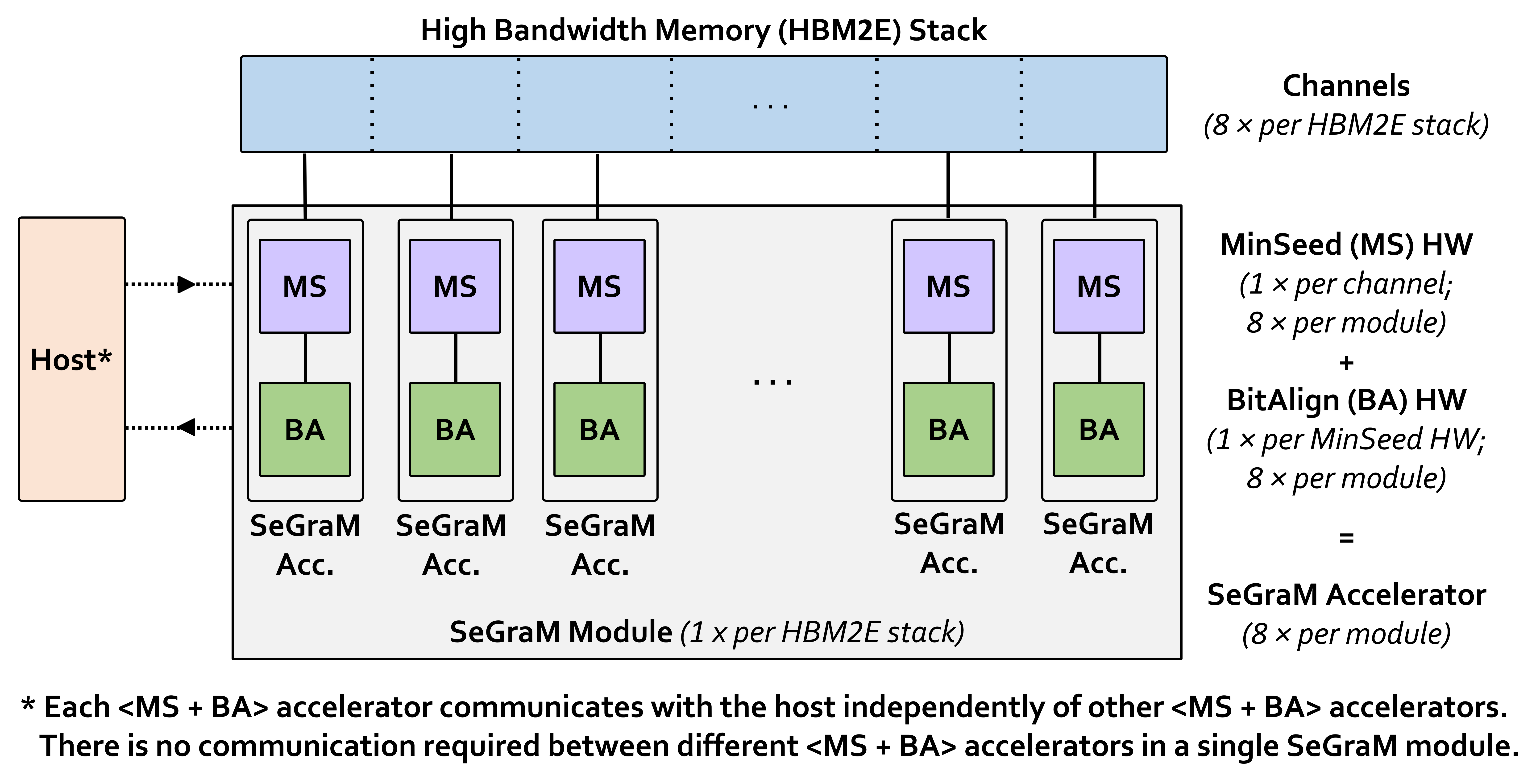}
\vspace{-18pt}
\caption{\reviscaIII{\fontsize{8.7}{8.7} \selectfont Overall system design of a \mech module. Our full design has four \mech modules and four HBM2E stacks.}} \label{fig:overall-system}
\vspace{-6pt}
\end{figure}

When the host transfers a single query read to \mech, the read is buffered before being processed. We employ \sgc{a} double buffering technique to hide the transfer latency of the query reads. Thus, this transfer is not on the critical path of \reviscaII{SeGraM’s} execution, since the next query read to be processed \sgc{has already been transferred and stored in the \emph{read scratchpad} of the \reviscaIII{\ms accelerator} when the current} query read is being processed. This ensures that the \reviscaII{host-to-\mech} connection is not a bottleneck and not on the critical path of the full \mech execution.}

\reviscaIII{Our \sgc{hardware} platform \sgc{includes} four \revISCA{off-chip} HBM2E stacks~\cite{hbm}, each with \sgc{eight memory} channels. Next to each HBM2E stack, we place one \mech module. A single \sgc{\mech module} consists of 8 \mech accelerators, where each accelerator is a combination of one \ms accelerator and one \ba accelerator. Each \sgc{\mech accelerator} has exclusive access to one HBM2E channel \sgc{to ensure} \damlaISCA{low-latency \sgd{and high-bandwidth memory access}}~\cite{ghose2019demystifying}, without \sgd{any interference} from other \mech accelerators in each module.\footnote{\reviscaIII{There is no communication required between different \mech accelerators in a single \mech module, and each \mech accelerator communicates with the host independently of other \mech accelerators.}}} By placing \mech in the same package as the four HBM2E stacks, we mimic the configuration of current commercial devices such as GPUs~\cite{v100,a100} and \reviscaIII{FPGA boards~\cite{vu37p,ad9h7,singh2020nero,singh2021fpga}}.

We replicate the graph-based reference and \sgc{hash-table-based index across all 4 independent HBM2E} stacks, \reviscaIII{which enables us to have 32 independent \mech accelerators running in parallel}. Within each stack, \reviscaIII{to balance the memory footprint across all channels,} we distribute the graph and index structures of all chromosomes (1--22, X, Y) based on their sizes \sgc{across the eight} independent channels. \reviscaI{For the datasets we use, the graph and index structures require a total memory of \SI{11.2}{\giga\byte} \reviscaII{per stack}, which is well \sgc{within} the capacity of \sgc{\sgd{a single} HBM2E stack \reviscaII{(i.e., \SI{16}{\giga\byte} \reviscaIII{in current technology})~\cite{hbm}} that} we include in our design.
}

\damla{
We design \reviscaIII{each \mech accelerator (\ms + \ba)} to operate in a pipelined fashion, such that we can hide the latency of \sgd{the \ms accelerator.}
\sgd{While} \ba is running, \reviscaIII{\ms finds the next set of minimizers, fetches the frequencies and seeds from the main memory}, and \sgd{writes} them to their associated scratchpads. In order to enable \sgd{pipelined execution of \ms and \ba}, 
we employ \sgd{the} double buffering technique for the \reviscaII{\emph{minimizer scratchpad} and \emph{seed scratchpad} \sgd{(as we do for the \emph{read scratchpad})}}.
}

\reviscaI{
If the minimizers \sgc{do not} fit in the minimizer scratchpad, we can perform a \sgd{batching} approach, where not all of the minimizers will be found and stored together. \sgc{Instead, a batch (i.e., a subset) of minimizers is found, stored, and used}, and then the next batch will be generated out of the read. \sgc{A similar optimization can be applied to the seed scratchpad if capacity issues arise}.
}
\vspace{-4pt}
\section{Use Cases of {\mech}} \label{sec:framework}

\sgc{As a result} of the flexibility and \sgd{modularity} of the \mech framework, we can run each accelerator (i.e., \ms and \ba) \reviscaIV{together for end-to-end \reviscaV{mapping} execution, or} separately.
\reviscaII{Thus, we describe three use cases of \mech: (1)~end-to-end 
mapping, (2)~alignment, and (3)~seeding.}

\vspace{1pt}
\reviscaII{\textbf{End-to-End Mapping.}}
For sequence-to-graph mapping, the whole \mech design \reviscaIV{(\ms + \ba)} should be \sgd{employed}, since both seeding and alignment steps are required \reviscaIV{(see Section~\ref{sec:background-mapping})}. 
\sgc{With} the help of the divide-and-conquer approach \sgc{inherited} from the \sgd{GenASM~\cite{cali2020genasm} algorithm},
we can use \sgc{\mech to perform} se\-quence-to-graph mapping for both short \sgc{reads} and long reads. \reviscaI{\sgc{Because} traditional sequence-to-sequence mapping is a special and simpler variant of sequence-to-graph mapping \sgc{(i.e., a graph where each node has an outgoing edge to exactly one other node)}, \mech can be used for \sgc{sequence-to-sequence mapping as well.}}

\vspace{1pt}
\reviscaII{\textbf{Alignment.}}
Since \sgc{\ba takes in} \reviscaIII{a graph-based reference and a query read as its inputs,} 
it can \sgd{be used} as a \sgd{standalone} sequence-to-graph aligner, without \sgd{\ms}. \reviscaI{Similar to \mech, \ba can also be used for sequence-to-sequence alignment, as sequence-to-sequence alignment is a special and simpler variant of sequence-to-graph alignment. \ba is orthogonal to \sgc{and can} be coupled with any seeding \sgd{(or filtering)} tool/accelerator.}

\vspace{1pt}
\textbf{Seeding.}
Similarly, \ms can be used \sgc{without \ba} as \sgc{a standalone seeding accelerator} for both graph-based mapping and \sgc{traditional} linear mapping. \ms is orthogonal to \sgc{and can} be coupled with any alignment tool or accelerator.
\vspace{-4pt}
\section{Evaluation Methodology} \label{sec:methodology}

\hspace{\parindent}
\textbf{Performance, Area and Power Analysis.}
We synthesize \revonur{and place \& route} the \ms and \ba accelerator datapaths using \revonur{the} Synopsys Design Compiler~\cite{synopsysdc} with a typical \sgd{\SI{28}{\nano\meter}} \revonurcan{low-power} process~\reviscaII{\cite{28nm}}.
Our synthesis targets \revIV{post-routing} timing closure at \hl{\SI{1}{\giga\hertz}} clock frequency.
\revISCA{Our power analysis for \sgd{CPU software baselines includes the power consumption of the CPU} socket and the dynamic DRAM power. In our power analysis for SeGraM, we include the dynamic HBM power~\cite{o2017fine,kim2021signal} and the power consumption of all logic and scratchpad/SRAM units.}
We then use \revonur{an in-house cycle-accurate} simulator and a \revonur{spreadsheet-based} analytical model parameterized with the synthesis and memory \revISCA{estimates} to drive the performance analysis.

\vspace{2pt}
\textbf{Baseline \sgd{Comparison Points}.}
\lijoel{First, we} compare \mech with two state-of-the-art \lijoel{CPU-based} sequence-to-graph mappers:  GraphAligner~\cite{rautiainen2020graphaligner} \sgc{and vg~\cite{garrison2018variation}}, running on \reviscaIII{an} Intel\textsuperscript{\textregistered} Xeon\textsuperscript{\textregistered} E5-2630 v4 CPU~\cite{intel_cpu} \reviscaIII{with 20 physical cores/40 logical cores with hyper-threading~\reviscaIV{\cite{magro2002hyper,koufaty2003hyperthreading,tullsen1995simultaneous,tullsen1996exploiting}}}, operating at \SI{2.20}{\giga\hertz}, with \SI{128}{\giga\byte} DDR4 memory. \damla{We run both \sgd{GraphAligner and vg} with 40 threads.}
We measure the execution time and power consumption \sgd{(using Intel's PCM power utility~\cite{intelpcm})} of 
\sgd{each CPU software baseline.} \lijoel{Second, we compare \mech with a state-of-the-art GPU-based sequence-to-graph-mapper, HGA~\cite{feng2021hga},\footnote{It is important to note that (1)~HGA does not support traceback and reports only the alignment score; and (2)~even though HGA is presented as a \reviscaII{sequence-to-graph alignment} 
\reviscaIII{tool} by its authors, we use it \reviscaII{as a sequence-to-graph mapping 
\reviscaIII{tool} since HGA takes all of the nodes of a given graph into consideration instead of a \reviscaIII{small} region of the graph. Thus, we compare HGA with \mech, which takes the complete graph as its input, instead of \ba, which takes a \reviscaIII{small} region of the graph \reviscaIV(i.e., subgraph) as its input.}} running on an NVIDIA\textsuperscript{\textregistered} GeForce\textsuperscript{\textregistered} RTX~2080~Ti~\cite{nvidia-rtx28080ti}. We measure the execution time of the GPU kernel only, ignoring any CPU overheads. We measure the power consumed by the entire GPU using the NVIDIA-smi tool~\cite{nvidia-smi} \revISCA{and subtract the static power to find the dynamic power}.} \lijoel{Third, we} compare \ba with a state-of-the-art \sgd{software-based} sequence-to-graph aligner, PaSGAL~\cite{jain2019accelerating}, and also with three state-of-the-art \sgd{hardware-based} sequence-to-sequence aligners: Darwin~\cite{turakhia2018darwin}, GenAx~\cite{fujiki2018genax}, and GenASM~\cite{cali2020genasm}. For these four baselines, we use the numbers reported by the papers. 

\vspace{2pt}
\textbf{Datasets.}
\label{sec:gengraph-methodology:datasets}
We evaluate \mech using the latest major release of the human genome assembly, GRCh38 \cite{ncbi38genome}, as the starting reference genome. \reviscaIII{To incorporate known genetic variations and thus form a genome graph}, we use 7 VCF files for HG001-007 from the GIAB project (v3.3.2)~\cite{giabvcf}. \revISCA{\reviscaIII{Across} the 24 graphs generated (one for each chromosome; 1--22, X, Y), in total, we have \SI{20.4}{\mega\nothing} nodes, \SI{27.9}{\mega\nothing} edges, \SI{3.1}{B\nothing} sequence characters, and \SI{7.1}{\mega\nothing} variations.}

\sgc{For} the read datasets, we generate four sets of long reads \revV{(i.e., PacBio and ONT datasets)} using PBSIM2~\cite{ono2021pbsim2} and three sets of short reads \revV{(i.e., Illumina datasets)} using Mason~\cite{holtgrewe2010mason}.
For the PacBio and ONT datasets, we have reads of length \SI{10}{\kilo\basepair}, \revonur{each simulated with 5\% and 10\% error rates}. The Illumina datasets have \hl{reads of length \SI{100}{\basepair}, \SI{150}{\basepair}, and \SI{250}{\basepair}}, \revonur{each simulated with a 1\% error rate}. \revISCA{Each dataset has 10,000 reads.}

\reviscaII{
\sgc{For our comparison with \sgd{HGA~\cite{feng2021hga}}, we follow the methodology presented in \sgd{\cite{feng2021hga}}, where we use the Breast Cancer Gene1 (BRCA1) graph~\cite{brca1_hgadataset} and three different read datasets simulated from the BRCA1 graph (using the \texttt{simulate} command from vg): R1 (\SI{128}{\basepair} $\times$ 278,528 reads), R2 (\SI{1024}{\basepair} $\times$ 34,816 reads), and R3 (\SI{8192}{\basepair} $\times$ 4,352 reads).}
}

\vspace{-4pt}
\section{Results} 
\label{sec:results}

\subsection{Area and Power Analysis}
\label{sec:gengraph-results:area-power}

Table \ref{table:gengraph-areapower} shows the area and power breakdown of the compute (i.e., logic) units, \damla{the scratchpads}, \revISCA{and HBM stacks} in \mech, \revonur{and the total area overhead and power consumption of (1)~a single \reviscaIII{\mech accelerator} (attached to a single channel), 
and (2)~32 \reviscaIII{\mech accelerators} (\sgc{with each accelerator attached to its own channel, across four HBM stacks that each have eight channels}). \sgc{The \reviscaIII{\mech accelerators}} \revonur{operate} at \SI{1}{\giga\hertz}.}

\damla{
\sgc{For a single \reviscaIII{\mech accelerator}, the area overhead is \SI{0.867}{\milli\meter\squared},
and the power consumption is \revISCA{\SI{758}{\milli\watt}}.}
\sgc{We find that the main contributors for the area overhead and power consumption are (1)~\sgc{the hop queue registers, which} constitute more than 60\% of the area and power of \reviscaIII{BitAlign's edit distance calculation logic; and (2)~the bitvector scratchpads.}}
\sgc{For \sgc{32~\reviscaIII{\mech accelerators}}, the total area overhead is}
\revISCA{\SI{27.7}{\milli\meter\squared}},
\sgc{with a power consumption of}
\revISCA{\SI{24.3}{\watt}}. \revISCA{When we add the HBM power to the power consumption of \sgc{32 \reviscaIII{\mech accelerators}}, the total power consumption becomes \SI{28.1}{\watt}.}} 

\reviscaIII{We conclude that \mech is very efficient in terms of both power consumption and area: a single \mech accelerator requires 0.02\% of area and 0.5\% of power consumption of \reviscaIII{an entire} high-end Intel processor~\cite{intel_cascadelake}.} 

\begin{table}[h!]
\vspace{-4pt}
\centering
\caption{\revISCA{Area and power breakdown of \mech.}}
\label{table:gengraph-areapower}
\vspace{-6pt}
\includegraphics[width=\columnwidth,keepaspectratio]{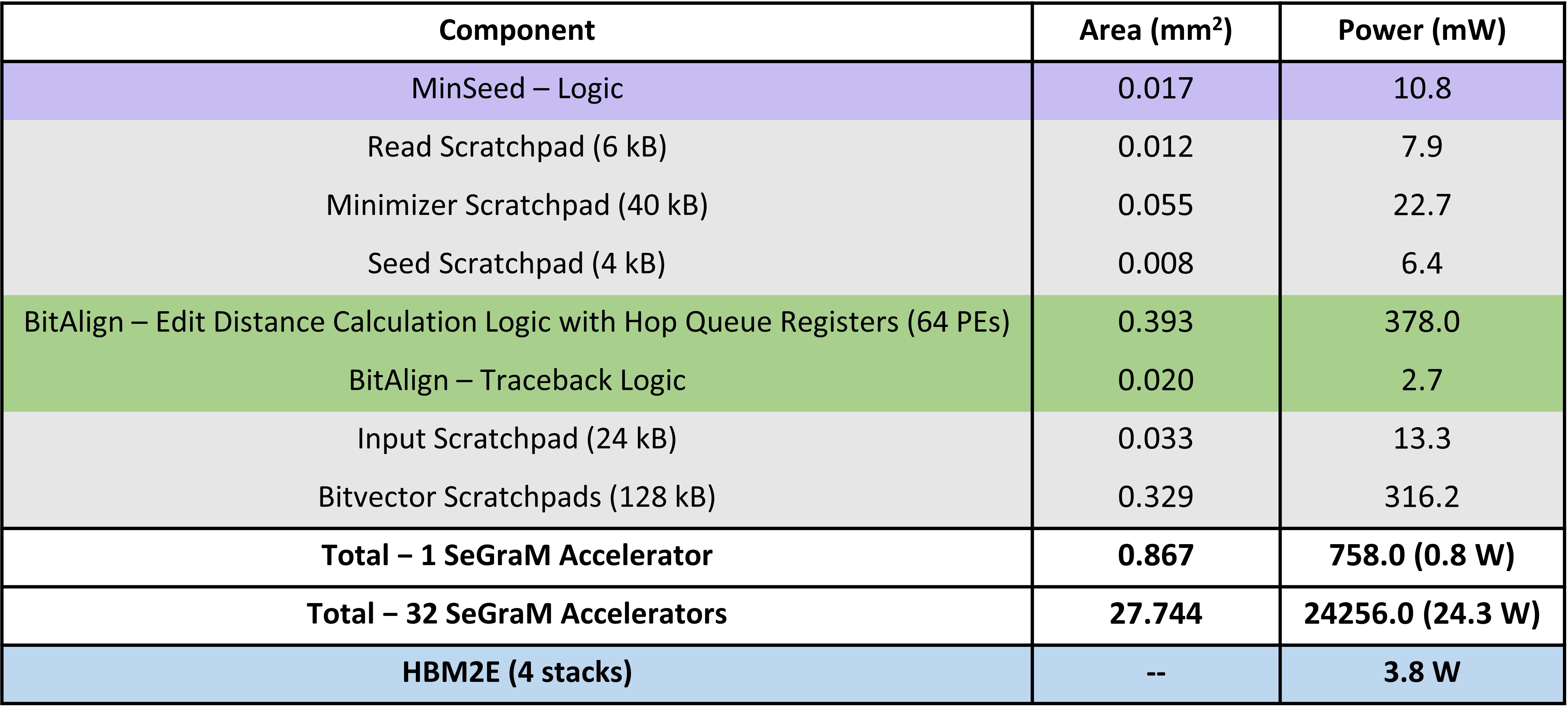}
\vspace{-20pt}
\end{table}

\subsection{Analysis of End-to-End \mech Execution}
\label{sec:results-segram}

\damla{
\hspace{\parindent}\textbf{\sgc{Comparison With} CPU \sgc{Software}.}
We compare end-to-end execution of \mech with two state-of-the-art sequence-to-graph mapping 
\reviscaIII{tools}, \sgc{GraphAligner~\cite{rautiainen2020graphaligner} and vg~\cite{garrison2018variation}. We evaluate 40-thread instances of GraphAligner and vg on the CPU.
We evaluate both 
\reviscaIII{tools} and \mech for both long reads and short reads.}}

\begin{figure}[ht!]
\centering
\includegraphics[width=\columnwidth,keepaspectratio]{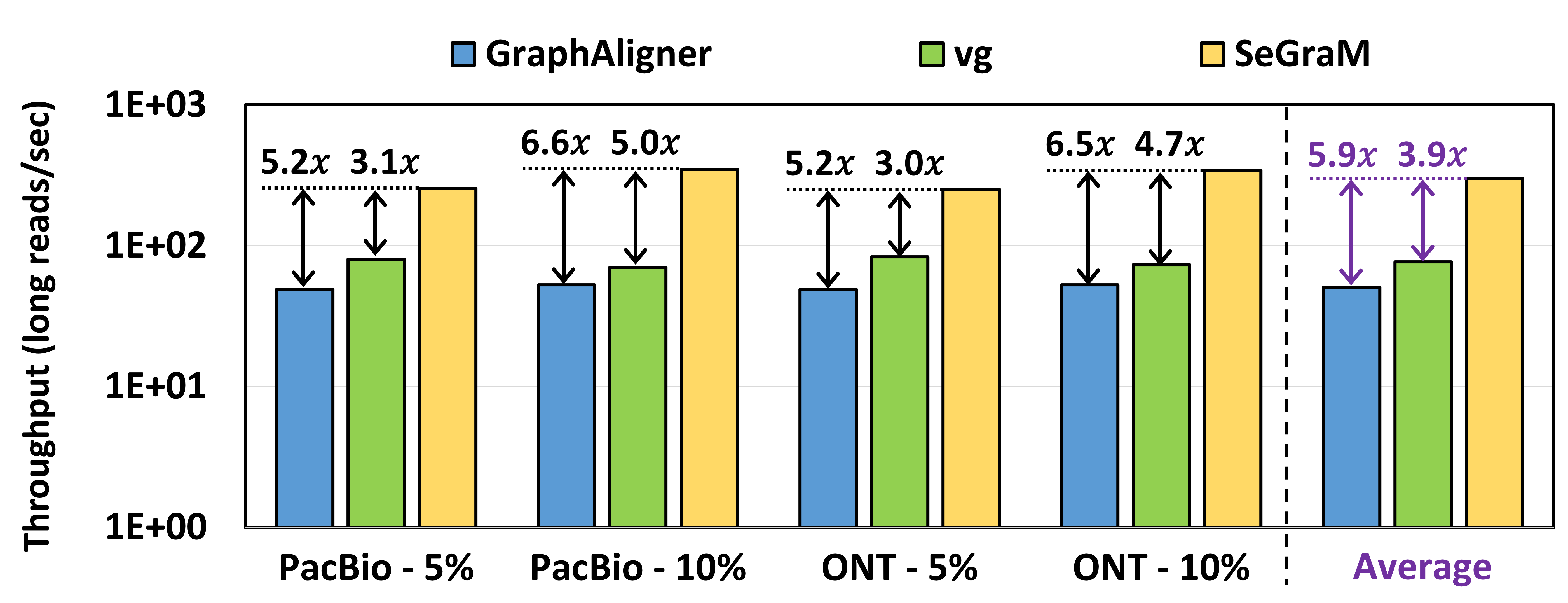}
\vspace{-18pt}
\caption{\sgc{Throughput of GraphAligner, vg, and \mech for long reads.} \reviscaII{5\%/10\% stands for PacBio/ONT datasets with 5\%/10\% read error rate.}} \label{fig:gengraph-throughput-result-long}
\vspace{-8pt}
\end{figure}

\reviscaI{Figure~\ref{fig:gengraph-throughput-result-long} shows the read mapping throughput (reads/sec) of \sgc{GraphAligner, vg, and \mech} when aligning long noisy PacBio and ONT reads against the graph-based representation of the human reference genome.}
We \sgd{make two observations. First,} on average, \mech provides \sgc{a throughput improvement of \revISCA{$5.9\times$} and $3.9\times$ \sgd{over GraphAligner} and vg, respectively}.
\sgc{We perform a power analysis (not shown), and find that \sgd{GraphAligner and vg consume \SI{115}{\watt} and \SI{124}{\watt}, respectively};
\mech reduces the power consumption for our long read datasets by $4.1\times$ and $4.4\times$, respectively.}
\sgc{Second, the throughput improvements do not change \sgd{greatly} with the \sgd{query read error rate}.
When we \sgd{examine} a single \mech execution, it takes \SI{35.9}{\micro\second} at a 5\% error rate, whereas it takes \SI{37.5}{\micro\second} at a 10\% error rate. \reviscaIII{However, we also find that} the total number of seeds \sgc{that need to be aligned} can vary based on the dataset\sgc{: for} the datasets we use in our analysis, \ms generates fewer seeds to align for the \sgc{10\%-error-rate} datasets compared to the \sgc{5\%-error-rate} datasets, \reviscaIII{thus,} \sgd{effectively canceling out the impact of increased execution time \reviscaIII{from 5\%-error-rate datasets to 10\%-error-rate datasets}}. 
As a result, we do not observe \sgd{a large} difference in overall throughput between the two datasets.}

\begin{figure}[h!]
\centering
\vspace{-8pt}
\includegraphics[width=\columnwidth,keepaspectratio]{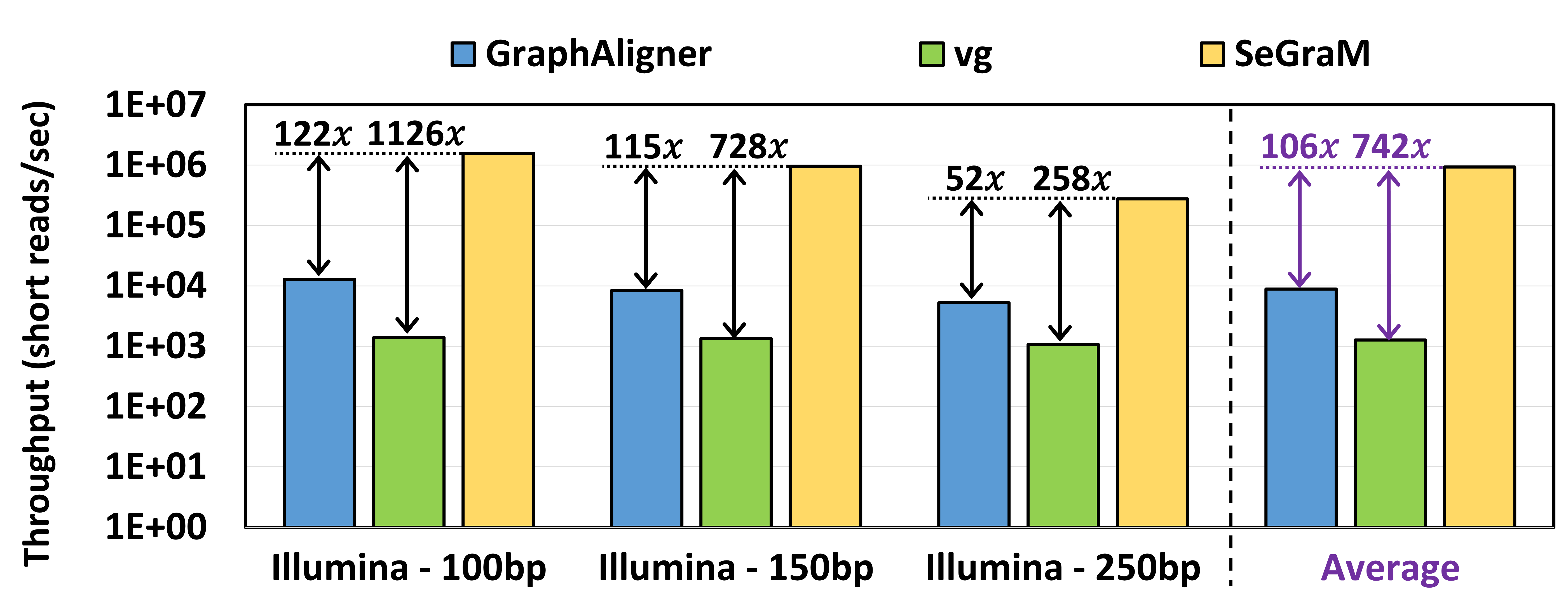}
\vspace{-18pt}
\caption{\sgc{Throughput of GraphAligner, vg and \mech for short reads.} \reviscaII{100/150/250bp stands for Illumina datasets with 100/150/250bp length reads.}} \label{fig:gengraph-throughput-result-short}
\vspace{-6pt}
\end{figure}

\reviscaI{Figure~\ref{fig:gengraph-throughput-result-short} shows the read mapping throughput of \sgc{GraphAligner, vg, and \mech} when aligning short \reviscaIV{accurate} Illumina reads against the graph-based representation of the human reference genome.} 
We \sgc{make two \sgd{observations}. First,} on average, \mech provides \sgc{a throughput improvement of \revISCA{$106\times$} and $742\times$ \sgd{over GraphAligner} and vg, respectively}.
\sgc{We perform a power analysis (not shown), and find that GraphAligner \sgd{and vg consume \SI{85}{\watt} and \SI{91}{\watt}, respectively};
\mech reduces the power consumption for our \reviscaIV{short} read datasets by $3.0\times$ and $3.2\times$, respectively.}
\sgc{Second, the throughput improvement of all three read mappers decreases as the read length increases.
This is because as the read length increases, 
the number of seeds to align increases as well, resulting in an increase in execution time.} \reviscaIII{However, \mech is affected by the increase in the number of seeds to align more than the baseline tools. Thus, SeGraM's throughput improvement over the baseline tools decreases when the read length increases \reviscaIV{(but still stays above $52\times$)}.}

\textbf{\sgc{Comparison With GPU Software}.}
We also compare end-to-end execution of \mech with the state-of-the-art GPU-based sequence-to-graph mapping 
\reviscaIII{tool}, HGA\sgc{~\cite{feng2021hga}}.
\sg{We} find that \mech provides 523$\times$, 85$\times$, and 17$\times$ \sgc{higher} throughput for the BRCA1-R1, BRCA1-R2, and BRCA1-R3 datasets \reviscaII{(Section~\ref{sec:gengraph-methodology:datasets})} over \sgc{HGA{, while reducing power consumption by $2.2\times$, $2.1\times$, and 1.9$\times$, respectively.}} 

\vspace{2pt}
\textbf{Sources of Improvement.}
\sgc{We identify four key reasons why \sgd{\mech achieves} such large performance improvements over GraphAligner, vg, and HGA.

First, we address \reviscaIV{the high cache miss rate bottleneck} by carefully designing and sizing the on-chip scratchpads (using empirical data) and the hop queue registers (which allow us to fetch all bitvectors for a hop within a single cycle). As a result, \sgd{\mech matches} the rate of computation for the logic units with memory bandwidth and \reviscaIV{memory} capacity. \sgd{Doing so} overcomes the high cache miss rates experienced by \reviscaII{GraphAligner}, and leads to a \sgd{large} reduction in the memory bandwidth that \mech needs to consume.

Second, we address \reviscaV{the} DRAM latency \reviscaIII{bottleneck} by taking advantage of the natural channel subdivision exposed by HBM. \reviscaIV{As mentioned} in Section~\ref{sec:gengraph-results:area-power}, we dedicate one \reviscaIII{\mech accelerator} per HBM channel. While \mech does not need to be implemented alongside an HBM-based memory subsystem, the multiple independent channels available in a single HBM stack allow us to increase \reviscaIII{accelerator-level} parallelism without introducing \reviscaIV{memory} interference among the \reviscaIII{accelerators}. Unlike with CPU/GPU threads, which share all of the memory channels, our channel-based isolation eliminates any \reviscaIV{inter-accelerator} interference-related latency in the memory system~\cite{muralidhara2011reducing}.

Third, our co-design approach for both seeding and alignment yields multiple benefits:
(1)~We make use of efficient and hardware-friendly algorithms for seeding and for alignment.
(2)~We eliminate the data transfer bottleneck between the seeding and alignment steps of the \reviscaIII{genome sequence analysis pipeline}, by placing their individual accelerators (\ms and \ba) adjacent to each other.
(3)~Our pipelining of the two accelerators within a \reviscaIV{\mech accelerator} allows us to completely hide the latency of \ms.%
}

\sgc{Fourth, while the performance of the software 
\reviscaIII{tools} scales sublinearly with the thread count, \mech scales linearly across three dimensions:}
%
\reviscaI{(1)~Within a single \sgc{\ba accelerator, by incorporating processing elements (PEs), we parallelize multiple \ba iterations.
This scales linearly up to the number of \sgd{bits processed} (128 in our case), as we can partition the iterations of the inner loop of the \ba algorithm (Lines 16--24 in Algorithm~\ref{bitalign-dc-alg}) across all of the PEs (i.e., we can incorporate as many as \reviscaIII{64~PEs} and still attain linear performance improvements).}
(2)~Since a single read is composed of multiple seeds/minimizers, we \sgd{use} pipelined execution (with the help of our double buffering approach) to execute multiple seeds in parallel.
(3)~With the help of multiple HBM stacks that \sgc{each contain the same content}, \sgd{we process} multiple reads concurrently \sgc{without introducing \reviscaIII{inter-accelerator} memory interference.}
\reviscaIII{Seed-level parallelism and read-level parallelism} can scale near-linearly as long as the memory bandwidth remains unsaturated, since \reviscaV{(1)~different seeds in a single read are independent of each other, (2)~different reads are independent of each other, and (3)~the memory bandwidth requirement \sgc{of} each read is low (\SI{3.4}{\giga\byte\per\second}).}
%
}

\sgc{Overall, we conclude that \mech provides substantial throughput improvements and \reviscaIII{power} savings over state-of-the-art software 
\reviscaIII{tools}, \sgd{for both long and short reads}.}

\damla{
\subsection{Analysis of \ba}

\hspace{\parindent}\textbf{Sequence-to-Graph Alignment.}
As we explain in Section~\ref{sec:framework}, \sgc{\ba can be used as a standalone accelerator} for sequence-to-graph alignment.
We compare \ba with the state-of-the-art \sgd{AVX-512~\cite{intelavx512}} based sequence-to-graph alignment 
\reviscaIII{tool}, PaSGAL~\cite{jain2019accelerating}. PaSGAL is composed of three main steps: (1)~DP-fwd, where the input graph and query read are aligned using the DP-based graph alignment approach to compute the ending position of the alignment;
(2)~DP-rev, where \sgc{the} graph and query \sgd{read} are aligned in the reverse direction to compute the starting position of the alignment;
and 
(3)~Traceback, \sgd{where,} using the starting and ending positions of the alignment, the corresponding section of the score matrix is re-calculated and traceback is performed to find the optimal alignment. 

Since the input of \ba is the subgraph and the query read, not the complete input graph, we compare \sgc{\ba \emph{only}} with the third step of PaSGAL for a fair comparison. \sgc{Figure~\ref{fig:gengraph-throughput-result-pasgal} shows the execution time of PaSGAL (using numbers reported in the PaSGAL paper~\cite{jain2019accelerating}) and \mech for \sgd{both short read (LRC-L1, MHC1-M1) and long read (LRC-L2, MHC1-M2)} datasets. We observe from the figure that} 
\mech provides $41\times$--$539\times$ speedup over the 48-thread \sgd{AVX-512 supported} execution of PaSGAL. 

\begin{figure}[ht!]
\centering
\vspace{-6pt}
\includegraphics[width=\columnwidth,keepaspectratio]{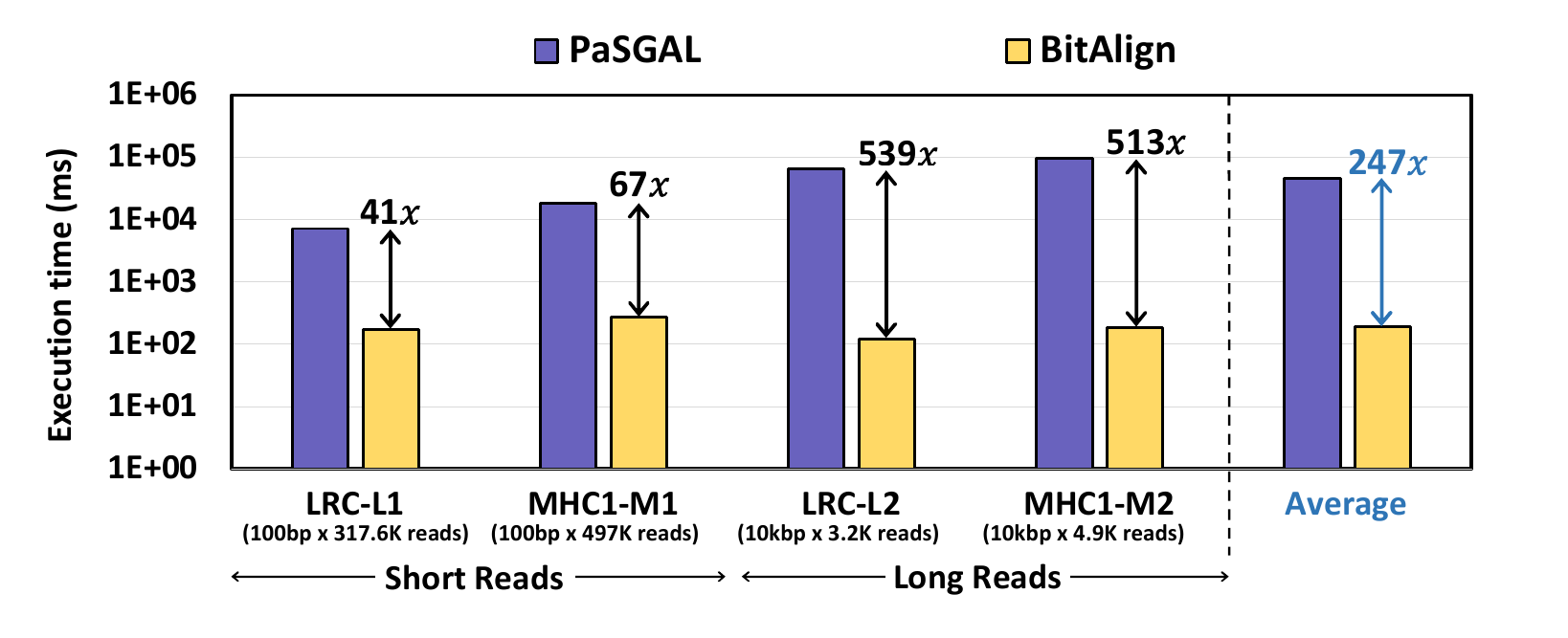}
\vspace{-22pt}
\caption{\revonur{Performance comparison of \sgc{PaSGAL and \sgd{\ba}} for sequence-to-graph alignment.}} \label{fig:gengraph-throughput-result-pasgal}
\vspace{-6pt}
\end{figure}

\sgc{We observe from the figure that \ba's speedup over PaSGAL is notably higher for long reads (i.e., the reads in the LRC-L2 and MHC1-M2 datasets).}
\revISCA{This is due to the} divide-and-conquer approach that \ba follows. Instead of aligning the full subgraph and the query read, with the help of the windowing approach \reviscaII{(Section~\ref{sec:bitgraph_algo})}, \ba decreases the complexity of sequence-to-graph alignment, \sgc{\sgd{which allows \ba} to efficiently align} both short and long reads.

\vspace{2pt}
\textbf{Sequence-to-Sequence Alignment.}
%
\sgc{As we discuss in Section~\ref{sec:framework}, \ba can be used for sequence-to-sequence alignment, by treating a linear sequence as a special case of a graph.}
To show the \sgd{benefits} of \ba for this special use case,
we compare \ba with the state-of-the-art hardware accelerators for sequence-to-sequence alignment: \sgc{the GACT accelerator from Darwin~\cite{turakhia2018darwin}, the SillaX accelerator from} GenAx~\cite{fujiki2018genax}, and GenASM~\cite{cali2020genasm}. GACT is optimized for long reads, SillaX is optimized for short reads, and GenASM is optimized for both short and long reads. We use the \sgc{optimized} configuration of each accelerator, \sgc{as} reported in \sgd{the} corresponding papers. 

\reviscaV{
Based on our analysis, we find that, on average, \ba provides (1)~a throughput improvement of $4.8\times$ over GACT for long reads, while consuming $2.7\times$ more power and $1.5\times$ more area; (2)~a throughput improvement of $2.4\times$ over SillaX for short reads; and (3)~a throughput improvement of $1.2\times$ and $1.3\times$ over GenASM for long reads and short reads, respectively, while consuming $7.5\times$ more power and $2.6\times$ more area. \reviscaII{Since both \ba and GenASM \reviscaIII{have fixed power consumption and area overhead independent of the dataset,} their power and area ~comparisons are the same for any input dataset.}
}
}

\reviscaI{
\vspace{2pt}
\textbf{\ba vs. GenASM.} 
As we explain in Section~\ref{sec:bitgraph_algo}, 
\sgc{\ba is a modified version of GenASM. Specifically,}
we decrease the memory footprint of GenASM by $3\times$, which helps us to increase the \sgc{number of bits processed by each} PE from 64 (in GenASM) to 128 (in \ba) and better utilize the \reviscaIII{bitvector scratchpad} capacity. With \sgc{more bits} per PE, \reviscaIII{the number of windows} executed \reviscaII{(Section~\ref{sec:bitgraph_algo})} decreases, which \sgd{enables \ba to have speedup} compared to GenASM. For example, for a read of \reviscaIV{\SI{10}{\kilo\basepair} length}, each window execution of GenASM takes 169 cycles, whereas it takes 272 cycles for \ba. However, the number of windows required to consume \sgc{\SI{10}{\kilo\basepair} (the length of one read)} is 250 for GenASM (\sgc{as the window length is the same as the number of bits per PE, or 64}), whereas this number is 125 for \ba, whose window length is equal to 128. 
\sgc{Multiplying the number of windows by the cycle time per window, we find that \ba (\SI{34.0}{\kilo~cycles}) performs better than GenASM (\SI{42.3}{\kilo~cycles}) by 24\% \reviscaII{($1.2\times$)}.}
}

\subsection{Analysis of \ms}

As we explain in Section~\ref{sec:overall_hw}, with the help of our pipelined design, \ms is \emph{not} on the critical path of the overall \mech execution. However, since \reviscaIV{\ms finds} subgraphs using candidate seed locations and sends them to \ba for the final alignment, it plays a critical role \sgc{for} the overall sensitivity \reviscaIII{(i.e., the metric that measures the accuracy \reviscaIV{of} a seeding or filtering mechanism in keeping (not filtering out) the seeds that would lead to the optimal alignment)} of our approach.
\reviscaIII{\ms does \emph{not} decrease the sensitivity of the overall sequence-to-graph mapping compared to the baseline \sgd{software} 
\reviscaIII{tools} since both \ms and the baseline software 
\reviscaIII{tools} implement the \reviscaIV{\emph{same}} optimization of discarding the seeds that have higher frequency than the threshold.}

\vspace{2pt}
\textbf{\ms vs. Filtering Approaches.}
\reviscaIII{
\ms does \emph{not} implement a filtering mechanism.\footnote{\reviscaIV{\ms is orthogonal to any filtering tool or accelerator~\cite{Xin2013,xin2015shifted,alser2017gatekeeper,alser2017magnet,alser2019shouji,alser2020sneakysnake,singh2021fpga,kim2018grim,mansouri2022genstore,nag2019gencache,subramaniyan2020accelerating,huangfu2020nest,huangfu2019medal,laguna2020seed,cong2018smem++}. Employing a filtering approach as part of our design would increase SeGraM's performance and efficiency, a study we leave to future work.}}.
Even though this \reviscaIV{leads to} a higher number of subgraphs that \sgc{must be processed by the (expensive)} alignment step compared to approaches that perform filtering,} \sgc{\ba's high efficiency \reviscaIV{greatly} alleviates the alignment bottleneck that exists in the software 
\reviscaIII{tools}}, as we show in Section~\ref{sec:results-segram}. For example, for a long read dataset, \sgc{while GraphAligner decreases the number of seeds extended from \SI{77}{\mega\nothing} to \SI{48}{\kilo\nothing} with its filtering/chaining approaches, \ms only decreases the number of seeds to \SI{35}{\mega\nothing}, yet \mech (\ms and \ba together) still outperforms GraphAligner.}
Similarly, for a short read dataset, \sgd{even though} GraphAligner decreases the number of seeds extended from \SI{828}{\kilo\nothing} to \SI{11}{\kilo\nothing}, \sgc{\sgd{\mech significantly outperforms GraphAligner even though} \ms only decreases \sgd{the number of seeds} to \SI{375}{\kilo\nothing}.%
}
\vspace{5pt}
\section{Related Work} \label{sec:related_work}

To our knowledge, \sgd{this is} the first \sgd{work} to propose
(1)~a hardware acceleration framework for sequence-to-graph mapping (\mech),  
(2)~a hardware accelerator for minimizer-based seeding (\ms), and 
(3)~a hardware accelerator for sequence-to-graph alignment (\ba).
No prior work \sgd{studies} hardware design for \reviscaI{graph-based genome \reviscaIV{sequence} analysis.}

\reviscaI{
\textbf{Software Tools for Sequence-to-Graph Mapping.}
There are \sgd{several} tools available \sgc{that specialize} for sequence-to-graph mapping or alignment. Examples of sequence-to-graph mapping tools \sgc{include} GraphAligner~\cite{rautiainen2020graphaligner}, vg~\cite{garrison2018variation}, \sgc{HGA~\cite{feng2021hga}}, HISAT2~\cite{kim2019graph}, and minigraph~\cite{li2020design}. \sgc{Other tools, \reviscaII{such as PaSGAL~\cite{jain2019accelerating}, abPOA~\cite{gao2021abpoa}, ASta\-rix~\cite{ivanov2020astarix,ivanov2022fast}, and Vargas~\cite{darby2020vargas}} \reviscaIII{perform} sequence-to-graph alignment only, without an indexing or a} seeding step. 
\sgc{All of these approaches are software-only, and we show quantitatively in Section~\ref{sec:results} that our \reviscaII{algorithm/hardware} co-design \sgd{greatly outperforms four} state-of-the-art tools: GraphAligner, vg, HGA, and PaSGAL.}

\textbf{Hardware Accelerators for Genome Sequence Analysis.}
Existing hardware accelerators for genome sequence analysis focus on accelerating only the traditional \reviscaIII{sequence-to-sequence mapping} pipeline, and cannot support genome graphs as their inputs. For example, GenStore~\cite{mansouri2022genstore}, ERT~\cite{subramaniyan2020accelerating}, GenCache~\cite{nag2019gencache}, NEST~\cite{huangfu2020nest}, MEDAL~\cite{huangfu2019medal}, SaVI~\cite{laguna2020seed}, SMEM++~\cite{cong2018smem++}, \sgd{Shifted Hamming Distance~\cite{xin2015shifted},} GateKeeper~\cite{alser2017gatekeeper}, MAGNET~\cite{alser2017magnet}, Shouji~\cite{alser2019shouji}, and SneakySnake~\cite{alser2020sneakysnake,singh2021fpga} accelerate the seeding and/or filtering steps of sequence-to-sequence mapping. 

Darwin~\cite{turakhia2018darwin}, GenAx~\cite{fujiki2018genax}, GenASM~\cite{cali2020genasm}, SeedEx~\cite{fujiki2020seedex}, WFA-FPGA~\cite{haghi2021fpga}, GenieHD~\cite{kim2020geniehd}, GeNVoM~\cite{khatamifard2021genvom}, FPGASW~\cite{fei2018fpgasw}, SWI\-FOLD~\cite{rucci2018swifold}, and ASAP~\cite{banerjee2018asap} accelerate read alignment \sgd{for only a linear} reference genome \reviscaII{(sequence-to-sequence alignment)}.
These accelerators have no way to track the \reviscaII{hops that exist in a graph-based reference}, and cannot be easily modified to support \reviscaII{hops.} 
\mech builds upon \sgd{the} hardware components of GenASM, 
\sgd{with enhancements \reviscaIII{(i.e., \ba)} and new components (i.e., \ms) that efficiently support both \reviscaIII{(1)~sequence-to-graph mapping/alignment and (2)~sequence-to-sequence mapping/alignment}}
(\sgd{by treating the linear reference genome as a genome} graph where each node has only one outgoing edge).

\sgc{A number of works propose} processing-in-memory (PIM)~\reviscaIV{\cite{mutlu2020modern,mutlu2019processing,ghose2019processing}} based accelerators for genome sequence analysis, such as GRIM-Filter~\cite{kim2018grim}, RAPID~\cite{gupta2019rapid}, PIM-Aligner~\cite{angizi2020pimaligner}, RADAR~\cite{huangfu2018radar}, BWA-CRAM~\cite{chowdhury2020dna}, FindeR~\cite{zokaee2019finder}, AligneR~\cite{zokaee2018aligner}, \reviscaII{and FiltPIM~\cite{khalifa2020filtpim}}. \reviscaIII{Similar to non-PIM \reviscaII{se\-quence-to-se\-quence} alignment accelerators, these PIM accelerators are designed} for a \sgd{linear reference genome} only, and cannot support genome graphs.

\sgc{Aside from the aforementioned} read mapping and read alignment accelerators, there are other genomics accelerators that target different genomics problems, such as nanopore basecalling~\cite{dunn2021squigglefilter,lou2020helix}, genome assembly~\cite{angizi2020pim}, exact pattern matching~\cite{jiang2021exma,wu2021sieve}, and \sgc{the} GATK-based variant calling pipeline~\cite{ham2020genesis,lo2020algorithm,wu2019fpga}. \sgc{Our work is orthogonal to these accelerators.}
}

\vspace{5pt}
\section{Conclusion} \label{sec:conclusion}

\sgh{%
\sgd{We \reviscaIV{introduce} \mech}, \revISCA{the first \emph{universal genomic mapping \sgd{acceleration framework}} that can support both sequence-to-graph and sequence-to-sequence mapping, for both short and long reads.}
\mech consists of co-designed algorithms} and accelerators for memory-efficient minimizer-based seeding \reviscaII{(\ms)} and highly-parallel bit\-vec\-tor-based sequence-to-graph alignment \reviscaII{(\ba)}, \reviscaIII{with inherent and efficient processing support for genome graphs}. 
\reviscaIV{We show that (1)~\mech provides greatly higher throughput and lower power consumption on both short and long reads compared to state-of-the-art software tools for sequence-to-graph mapping, and (2)~\ba significantly outperforms a state-of-the-art sequence-to-graph alignment tool and three state-of-the-art hardware solutions that are specifically designed for sequence-to-sequence alignment.
We conclude that \mech is a promising framework for accelerating \reviscaIII{both graph-based and traditional} \reviscaIV{linear-sequence-based} genome sequence analysis. \reviscaIII{We hope that} our work inspires future \reviscaIV{research and design} efforts at accelerating \reviscaII{graph-based genome analysis via efficient algorithm/hardware co-design.}} 
\vspace{5pt}

\begin{acks}
\reviscaI{
We thank the anonymous reviewers of HPCA 2022 and ISCA 2022 for their feedback. We thank the SAFARI Research Group members for valuable feedback and the stimulating intellectual environment they provide. We acknowledge the generous gifts of our industrial partners, especially Google, Huawei, \sgd{Intel,} Microsoft, VMware, \reviscaIV{Xilinx}. This research was partially supported by the Semiconductor Research Corporation.
}
\end{acks}

{
\clearpage
\balance
\setstretch{1}
\bibliographystyle{IEEEtranS}
\bibliography{main}
}


\end{document}